\documentclass[12pt]{article}

\usepackage[utf8]{inputenc}
\usepackage[T1]{fontenc}
\usepackage{lmodern}
\DeclareUnicodeCharacter{2061}{} 

\usepackage{graphicx}
\usepackage{multirow}
\usepackage{amsmath,amssymb,amsfonts}
\usepackage{amsthm}
\usepackage{mathrsfs}
\usepackage[title]{appendix}
\usepackage{xcolor}
\usepackage{textcomp}
\usepackage{manyfoot}
\usepackage{booktabs}
\usepackage{algorithm}
\usepackage{algorithmicx}
\usepackage{algpseudocode}
\usepackage{listings}
\usepackage{geometry}
\usepackage{caption}
\usepackage{subcaption}
\usepackage[round]{natbib}
\usepackage{pdflscape}
\usepackage{booktabs}
\usepackage{caption}
\usepackage{threeparttable}
\usepackage{adjustbox}   
\usepackage{makecell}
\usepackage{breakurl} 
\usepackage[hyphens]{url}
\usepackage{hyperref} 

\title{Evaluating Ethnic Income Gap in China: The Case of Han, Mongol, and Manchu in Liaoning and Inner Mongolia}
\author{Xinyan (Emily) Deng\thanks{Email: \texttt{xdeng20@jhu.edu}}}
\date{May 1, 2024}

\begin{document}
\maketitle

\begin{abstract}
This study analyzes the 2018 Chinese Household Income Project survey data to evaluate the income gaps between an “outsider” ethnic minority group, the Mongols, an “insider” ethnic minority group, the Manchus, and the majority Han group in urban and rural areas of Liaoning province and Inner Mongolia in China. Three statistical methods, a simple first-order OLS linear regression, linear regressions with interaction terms, and the Blinder-Oaxaca Decomposition, are used to investigate the income disparity amongst the three groups. The results indicate that Mongols suffer a significant ethnic wage penalty attributable to possible discrimination in the rural areas of these two provinces, while the urban income gaps between the three groups can mostly be explained by participation in public sector occupations or affiliation with the Chinese Communist Party (CCP). In rural settings, Mongols also have higher returns to public sector jobs and CCP membership compared to the other two ethnic groups. The findings suggest that Chinese affirmative actions regarding ethnic policy are effective in accelerating the integration of ethnic minorities with Han in the outcomes of the labor market. This conclusion is consistent with previous studies.
\end{abstract}

\maketitle

\newpage
\section*{Acknowledgement}{I would like to express my sincere gratitude towards Professor Robert Moffitt, my mentor, for generously providing guidance and insightful advice for me throughout the duration of this project, from the suggestion of data sources to the interpretation of the econometric models. His expertise in the field of economics has been instrumental in shaping the direction of my research and enhancing the overall quality of my work. I am deeply grateful for the time and effort that Professor Moffitt dedicated to providing me with constructive feedback, thoughtful insights, and encouragement at every stage of this project.}
\newpage

\section{Introduction}\label{sec1}

As the second most populated country in the world, China has a diverse population that consists of 56 ethnic groups, with Han being the majority group making up over 90\% of the population and 55 minority groups accounting for the rest. Historically, ethnic minority groups reside in less economically developed regions and are typically more involved in agricultural activities. Since the establishment of the People’s Republic of China in 1949, the Chinese government has developed a substantial amount of affirmative action to stimulated economic growth of each ethnic minority group while preserving their distinct culture. The preferential policies are comprehensive, covering aspects from local government to family policies with the goal of achieving the optimal national form of “diversity and unity within Chinese configuration” \citep{bib15}. For example, in regions where a minority group is concentrated, autonomous governments are established to ensure that minority interests are represented in local policies. In such regions, the head of the government is an ethnic minority, and the proportion of minority governmental officials matches the proportion of the minority population in the region. In addition, as specified in Article 22 of the Regional Ethnic Autonomy Law, “when employing cadres, autonomous agencies in ethnic autonomous areas should give proper consideration to hiring members of the nationality exercising autonomy and other minorities from the area” \citep{bib11}. The required preferential hiring policy of public agencies encourages higher employment rates for minorities in autonomous regions. In terms of education, the minority groups receive 10 to 30 extra points in Gaokao, the national university entrance exam, making higher education more accessible to minority students. Even the One Child Policy, which restricts each Han family to only have one child, does not apply to minority families \citep{bib12}. 

Despite attempts by the government to preserve cultural diversity, most minority groups still show assimilation trends in the largely Han-dominated environment. As a member of the Xibe minority group myself, I find it hard to connect to my heritage as I do not have a distinct appearance specific to my ethnicity nor am I sufficiently exposed to the culture. For many minorities like me, the status of being part of an ethnic minority group is merely represented by the classification on the identification card. In a previous study on ethnic income disparity in China, such minority groups in which members are largely assimilated and cannot be easily identified as different from the Han majority group are defined as “insider” minority groups. On the other hand, minority groups whose appearances, names, customs, and sometimes even first language diverge from those of Han are referred to as the “outsider” minority groups \citep{bib10}. In labor markets, due to the visible differences between the outsider groups from Han, the employers can easily distinguish between the two groups, causing bias to be a possible factor of influence in making hiring decisions, which may consequently engender differences in labor market outcomes for the outsider groups and other groups. While many studies in the past have attempted to explain the wage gap between Han and ethnic minority groups in different regions of China, very few have scrutinized the disparity among the minority groups because of their different attributes. How the outsider groups, the insider groups, and the majority group fare differently in China’s labor market remains an underexplored topic. The challenge in addressing this research question lies in the difficulties associated with choosing the appropriate ethnic groups to analyze. Considering the very different behaviors of different ethnic groups in the labor market, it would be hard to draw a unified conclusion for all “outsider” groups and all “insider” groups. Most “outsider” minority groups reside in clusters at their historical settlement areas, making spatial factors a confounding covariate complicating the relationship between ethnicity and income level. 

To answer this question, I examine whether income gaps exist between the Mongols, a major outsider group in northern China, the Manchus, an insider group, and the Han while controlling for covariates related to income level. The regional focus of this study is on Inner Mongolia and Liaoning, containing an autonomous region for Mongols. Located in the northeastern part of China, these two provinces take up approximately 13.8\% of China’s territorial area. Although not conventionally considered economically well-developed areas where the major economic activities are in heavy industry, agriculture, and animal husbandry, the two provinces have an average GDP per capita that is close to the GDP per capita of the whole country in 2018. By limiting the research to this specific region, I was able to evaluate the effects of affirmative actions in minority autonomous regions. In addition, Inner Mongolia is the region where the Mongolian culture is the most well-preserved, and assimilation levels of Mongols remain the lowest. Theoretically, under the preferential policies, Mongols should be more involved in governmental jobs. After accounting for the differences in other related factors, the Mongols as an outsider group should receive a lower wage due to ethnic discrimination while the Manchus should not have any significant difference in earnings compared to the Han. Using the Chinese Household Income Project (CHIP) survey data from 2018 and three different regression methods, I explore how much of the difference in the income levels of the three groups can be explained by the different average endowments such as gender, age, years of schooling, occupational field, etc. of the three groups. Three types of statistical models – simple OLS linear regressions, OLS linear regressions with interaction terms, and pairwise Blinder-Oaxaca decomposition – are used on a total of 6,437 observations for this study. Results from the three types of analyses all suggest that the Manchu group has similar labor market outcomes as the Han in both urban and rural areas while the Mongol group receives lower wages after accounting for the personal and occupational characteristics in the rural setting. Regression results also indicate that affiliation with the Communist Party and employment in the public sector are important factors explaining the ethnic gaps in urban regions.

The structure of this paper proceeds as follows: Section \ref{sec2} discusses previous studies on the ethnic income disparity in China. Section \ref{sec3} describes the data source used for the research. Section \ref{sec4} presents the statistical methodology and equations used. Section \ref{sec5} expands on the findings of the regression models. Section \ref{sec6} reviews the results and suggests possible areas of application of the findings.

\section{Literature Review}\label{sec2}

Although there is not a substantial body of research that address the ethnic income gap in China, some researchers have made prominent progress in attempting to understand the earnings disparity between different ethnic groups under specific regional context. The three most popular themes that existing literature explores are (1) the evolution of the ethnic wage gap over time, (2) the wage gap between Han and a specific minority group, and (3) the effect of a specific factor on the ethnic differences in labor market outcomes.

In the paper “Outsider ethnic minorities and Wage Determination in China” by \citet{bib10}, from which I adopted the concepts of “insider” and “outsider” minority groups, the researchers analyze 2011 CHES data containing 34 ethnic groups in the urban areas within seven provinces to determine the differences in how outsider and insider minority groups fare in Chinese labor markets. Although the researchers found that on average the minority groups do not experience different labor market outcomes after controlling for the relevant covariates, they still discovered that outsider minority groups suffer from earnings penalties while insider groups fare relatively similarly to Han. My paper seeks to add the complexity of regional effects to this study by limiting the regional focus to Liaoning and Inner Mongolia and adding a comparison between the rural and urban areas.

During the past few decades, China underwent dramatic economic developments. Therefore, many studies look at the changes in the labor market outcomes for ethnic minorities and Han during this era of drastic transformations in the economy. In an earlier study by \citet{bib4}, the authors use data from 1988 and 1995 surveys conducted by the Institute of Economics at the Chinese Academy of Social Sciences to examine how the rural ethnic income gap has changed during China’s transition into a market economy. The research found that although the income levels of both groups increased from 1988 to 1995, the minority-majority income gap almost doubled during the 7 years and that the driving force behind this widening gap is the difference in the geographical location of minority groups and the majority group. A very similar topic on the effects of steps towards a market economy is also elaborated by \citet{bib5}. Instead of merging the minorities into one category in the analysis, this paper investigates how the earnings gaps between Han and nine groups have been affected by the economic transition respectively. By studying a younger and an older cohort using the 2005 China Inter-Census Survey, the researchers find that Koreans enjoy an earnings premium compared to Han, and the premium increases over time; the younger cohorts of Uighur, Miao, and Zhuang suffered more wage penalty than the older group; there were no sign of premium nor penalty for Mongol, Hui, Manchu, and Tujia minorities. The study also finds that affirmative action by the government has caused Mongol, Tibetan, Uighur, and Tujia groups to have a higher probability of working in the public service sector. This is an improvement from the 2003 study in that it accounts for the within-minority variances.

\citet{bib9} investigated the changes in ethnic earnings gaps and poverty statuses outside of the 5 autonomous regions in China from 2002 to 2013 using CHIP survey data from the two years accordingly. Contrary to previous results, researchers found both the wage and poverty gaps narrowing during this studied decade. The researchers ascribe this discovery to the indifference to ethnic identity in the less-developed regions where minorities reside. Relatively recent research by \citet{bib6} examines a similar period using the same data sources CHIP from years 2002, 2013, and 2018. By analyzing the Yi and Manchu groups separately, researchers found that the average income per capita remained similar for Manchu and Han while the gap between Yi and Han narrowed from 2002 to 2018. This trend can be attributed to the increasing reliance on agricultural income of Yi.

Since the 55 ethnic minority groups exhibit large variations in the ways they perform in the labor markets, some researchers choose to focus on the earnings gap between Han and a specific minority group. Gustafsson and Ding (\citeyear{bib2}, \citeyear{bib3}) conducted two studies that scrutinized the case of the Hui minority group in the Ningxia autonomous region. Both studies utilize 2007 survey data conducted by the Ningxia Survey Team of the National Bureau of Statistics. While the first study looks broadly at the general earnings difference between Hui and Han within the rural Hui autonomous region, the second study inquires into the disparity in years of schooling for Hui and Han. Although the first research found no significant difference in average income per capita for Hui and Han in rural Ningxia, they found large discrepancies in the length of education between Hui and Han for both genders. Elaborated in the second paper, the researchers uncovered that for Hui minorities, education requires higher opportunity costs and provides a lower payoff compared to Han. As a result, Hui parents also spend fewer resources on their children's education. 

Aside from Hui, the Uyghur group in the Xinjiang autonomous region has also received a considerable amount of scholarly attention due to frequent ethnic unrest in the region. \citet{bib17} uses a subsample of the 2005 Urumchi survey to evaluate the Uyghur-Han earnings differential and observes a significant earnings gap in the non-state sector with no distinction in the state sector, reinforcing the fact that the affirmative policies in Xinjiang province are effective. \citet{bib16} use the 2005 Xinjiang Population mini-census data to decompose the ethnic earnings discrepancy between Uyghur and Han in Xinjiang into the effects of the rural-urban divide, occupational sector, migration, etc. This study arrives at a very similar conclusion as \citet{bib17} as it claims that the Uyghur-Han income difference is negligible in government jobs and is positively correlated with the marketization of the employment sector.
Another approach that researchers have taken to narrow down the focus is to look at the impacts of specific endowment variables. Educational background, being one of the most essential parts of human capital, is a popular covariate inspected by many studies. \citet{bib13} investigated the reasons behind the significantly lower wages for minority migrant workers and Han migrant workers in the Pearl River Delta region. They relied on surveys from 2008 to 2010 on migrant workers with the highest educational attainment of college or lower in the studied location. The regression results demonstrate that minorities in general have shorter years of schooling than Han; regardless of ethnicity, urban Hukou holders exhibit longer years of schooling than rural Hukou holders. Therefore, urban Hukou status helps decrease the inequality caused by differences in schooling for minorities, although only to a limited extent. This helps explain why minority migrants with an urban Hukou have a higher return in the labor market than Han migrants with an urban Hukou. \citet{bib1} also delve into the influence of education on the disparity in income between ethnic minorities and the Han. Analyzing CHNS data from the years 1993 to 2011, the authors discover that additional education indeed will help reduce the income gap by increasing minority income levels. 

Aside from length of education, \citet{bib7} assessed the impact of another key feature – immigration status on the income distribution among Han and minorities with the CHES 2012 dataset. He finds that migration in general leads to positive income gains for all households regardless of ethnicity although Han households benefit more from migrations from rural to urban areas. However, regression uncovers that migration not only widens the Han-minority divide but also intensifies the inequality among ethnic minority groups. In the very recent paper by \citet{bib14}, the researcher uses 2012 and 2017 Chinese General Social Survey datasets to explore how social capital affects Han and minority earnings. Data shows that as county-level social capital rises, its beneficial impact lessens for the income of minority groups whereas strengthens for Han. The difference in returns to social capital can be ascribed to the fact that the minorities’ accessed networks are on average lower in occupational prestige than Han’s.

Overall, although with different emphases, the existing papers predominantly conclude 5 main explanations for the divergence in earnings, which are location, human capital, occupational sector, cultural barrier, and discrimination. It can be concluded from past studies that after accounting for the first four explanatory factors, most “insider” groups do not differ much from the Han in terms of earnings levels while some “outsider” groups suffer from ethnic income penalties. While most preceding research narrows down the regional focus to either the urban population or the rural population, this study compares the different labor market outcomes in these two regions. 

\section{Data}\label{sec3}
This study uses a subsample of Liaoning and Inner Mongolia from the 2018 Chinese Household Income Project (CHIP) survey dataset. The CHIP2018 sample is a derivation of the extensive annual integration household survey conducted by the National Bureau of Statistics in 2018. The surveyed households are classified into urban and rural based on the area they reside regardless of their Hukou statuses. Two surveys are used to obtain the data, one for households in urban areas and one for households in rural areas, with slight distinctions in the sections for social relief programs and labor time allocation, which are both irrelevant to the purpose of this study. The dataset contains 11,506 urban household observations and 9,239 rural household observations from 15 provinces with the goal of capturing household and individual characteristics and income trends across the eastern, western, and central parts of China. The survey contains comprehensive financial as well as personal information about each household member, including gender, age, educational attainment, occupational sector, employment type, political affiliation, Hukou status, living situation, social insurance, annual income, household assets and debts, etc. After removing missing values and extracting observations from only the two provinces of interest of the three ethnic groups being examined, the final effective sample size of the study is 6,437.

\begin{landscape}
\begin{table}
\centering
\caption{Descriptive Summary Statistics}
\label{tab:1}
\resizebox{1.55\textwidth}{!}{%
\begin{threeparttable}
\setlength{\tabcolsep}{4pt}
\renewcommand{\arraystretch}{1.1}
\footnotesize
\begin{tabular}{lcccccccccccc}
\toprule
 &
  \multicolumn{4}{c}{\textbf{Urban}} &
  \multicolumn{4}{c}{\textbf{Rural}} &
  \multicolumn{4}{c}{\textbf{Pooled}} \\
  \cmidrule(lr){2-5}\cmidrule(lr){6-9}\cmidrule(lr){10-13}
 &
  \multicolumn{1}{c}{Han} &
  \multicolumn{1}{c}{Mongol} &
  \multicolumn{1}{c}{Manchu} &
  \multicolumn{1}{c}{\textbf{Total}} & 
  \multicolumn{1}{c}{Han} &
  \multicolumn{1}{c}{Mongol} &
  \multicolumn{1}{c}{Manchu} &
  \multicolumn{1}{c}{\textbf{Total}} &
  \multicolumn{1}{c}{Han} &
  \multicolumn{1}{c}{Mongol} &
  \multicolumn{1}{c}{Manchu} &
  \multicolumn{1}{c}{\textbf{Total}} \\
\midrule
n (\# of observations) &
  3234 &
  360 &
  269 &
  3863 &
  1785 &
  412 &
  377 &
  2574 &
  5019 &
  772 &
  646 &
  6437 \\
Average Age &
  40.39 &
  33.89 &
  37.54 &
  39.58 &
  44.29 &
  35.65 &
  42.6 &
  42.66 &
  41.78 &
  34.83 &
  40.49 &
  40.81 \\
Gender &
   &
   &
   &
   &
   &
   &
   &
   &
   &
   &
   &
   \\
Male &
  0.50 &
  0.51 &
  0.46 &
  0.50 &
  0.51 &
  0.48 &
  0.52 &
  0.51 &
  0.50 &
  0.49 &
  0.50 &
  0.50 \\
Female &
  0.50 &
  0.49 &
  0.54 &
  0.50 &
  0.49 &
  0.52 &
  0.48 &
  0.49 &
  0.50 &
  0.51 &
  0.50 &
  0.50 \\
Average Years of Schooling &
  10.43 &
  11.01 &
  10.22 &
  10.47 &
  7.53 &
  8.29 &
  7.64 &
  7.66 &
  9.39 &
  9.54 &
  8.68 &
  9.33 \\
Average Family Size &
  3.01 &
  3.07 &
  3.17 &
  3.03 &
  3.30 &
  3.77 &
  3.64 &
  3.43 &
  3.12 &
  3.45 &
  3.44 &
  3.19 \\
Public/Gov Job Proportion &
  0.32 &
  0.53 &
  0.19 &
  0.33 &
  0.12 &
  0.26 &
  0.13 &
  0.13 &
  0.27 &
  0.46 &
  0.16 &
  0.28 \\
Communist Proportion &
  0.14 &
  0.21 &
  0.10 &
  0.15 &
  0.05 &
  0.09 &
  0.05 &
  0.05 &
  0.11 &
  0.14 &
  0.07 &
  0.11 \\
Average Income\footnotemark[1]\footnotemark[2] &
  \begin{tabular}[c]{@{}c@{}}51459.32\\      (1222.21)\end{tabular} &
  \begin{tabular}[c]{@{}c@{}}60946.16\\      (3753.76)\end{tabular} &
  \begin{tabular}[c]{@{}c@{}}40286.40\\      (2778.43)\end{tabular} &
  \begin{tabular}[c]{@{}c@{}}51593.69\\      (1105.11)\end{tabular} &
  \begin{tabular}[c]{@{}c@{}}25496.63\\      (1037.38)\end{tabular} &
  \begin{tabular}[c]{@{}c@{}}26766.81\\      (3144.09)\end{tabular} &
  \begin{tabular}[c]{@{}c@{}}36311.78\\      (6766.41)\end{tabular} &
  \begin{tabular}[c]{@{}c@{}}27521.03\\      (144.57)\end{tabular} &
  \begin{tabular}[c]{@{}c@{}}45147.61\\      (988.02)\end{tabular} &
  \begin{tabular}[c]{@{}c@{}}51023.12\\      (2980.66)\end{tabular} &
  \begin{tabular}[c]{@{}c@{}}38283.44\\      (3672.80)\end{tabular} &
  \begin{tabular}[c]{@{}c@{}}45040.68\\      (918.77)\end{tabular} \\
Average Log Income &
  \begin{tabular}[c]{@{}c@{}}10.48\\      (0.03)\end{tabular} &
  \begin{tabular}[c]{@{}c@{}}10.70\\      (0.07)\end{tabular} &
  \begin{tabular}[c]{@{}c@{}}10.28\\      (0.08)\end{tabular} &
  \begin{tabular}[c]{@{}c@{}}10.48\\      (0.02)\end{tabular} &
  \begin{tabular}[c]{@{}c@{}}9.70\\      (0.05)\end{tabular} &
  \begin{tabular}[c]{@{}c@{}}9.52\\      (0.19)\end{tabular} &
  \begin{tabular}[c]{@{}c@{}}9.78\\      (0.10)\end{tabular} &
  \begin{tabular}[c]{@{}c@{}}9.70\\      (0.04)\end{tabular} &
  \begin{tabular}[c]{@{}c@{}}10.29\\      (0.02)\end{tabular} &
  \begin{tabular}[c]{@{}c@{}}10.36\\      (0.08)\end{tabular} &
  \begin{tabular}[c]{@{}c@{}}10.03\\      (0.07)\end{tabular} &
  \begin{tabular}[c]{@{}c@{}}10.27\\      (0.02)\end{tabular}\\
  \bottomrule
\end{tabular}
\begin{tablenotes}[flushleft]
\footnotesize
\item[1] Numbers in parentheses are standard deviations.
\item[2] Unit for income is CNY/year.
\end{tablenotes}
\end{threeparttable}
}
\end{table}
\end{landscape}

\begin{figure}
     \centering
     \begin{subfigure}[b]{15cm}
         \centering
         \includegraphics[width=15cm]{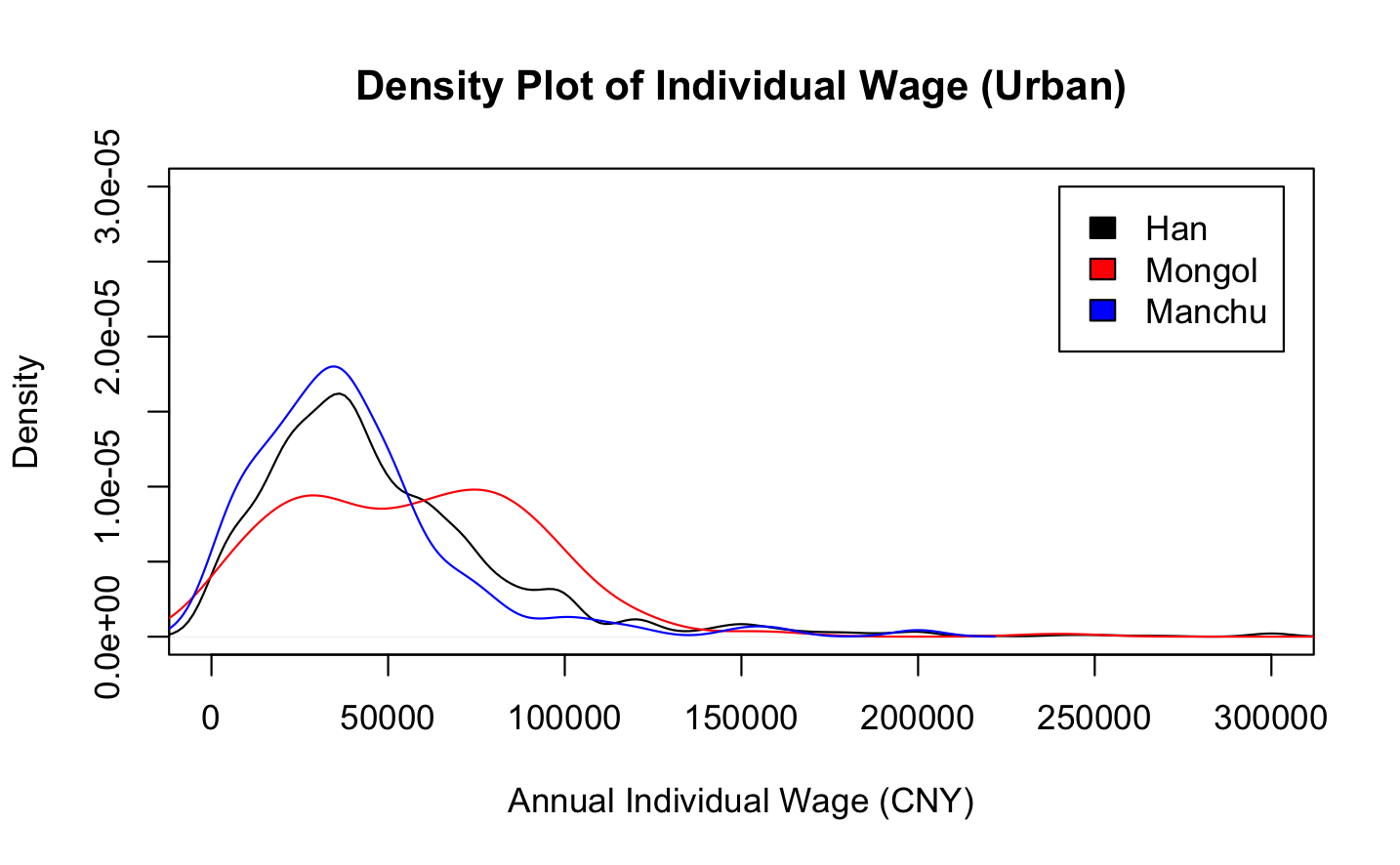}
         \caption{Income Distribution for the Three Groups in Urban Areas}
         \label{fig:1a}
     \end{subfigure}
     \vfill
     \begin{subfigure}[b]{15cm}
         \centering
         \includegraphics[width=15cm]{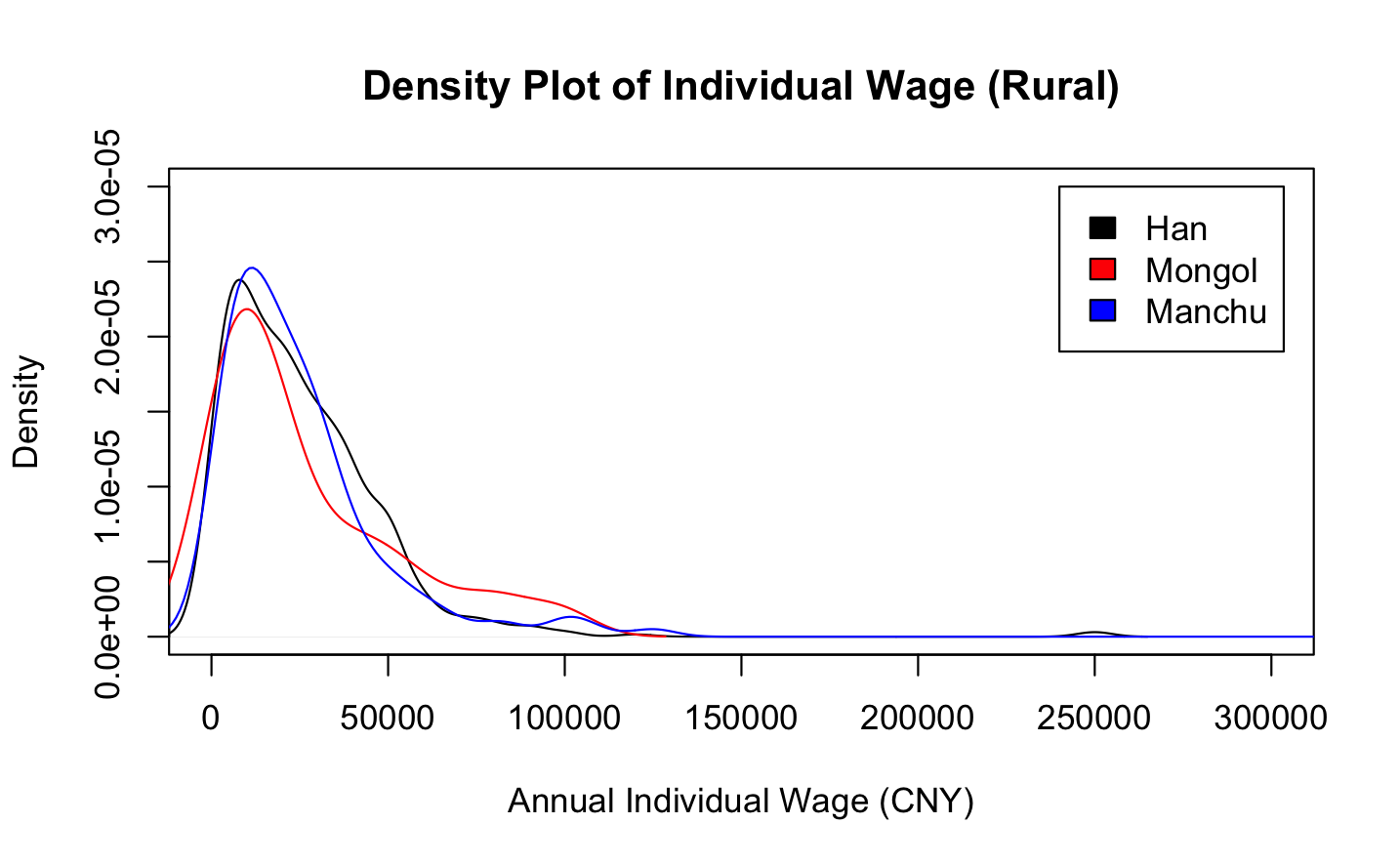}
         \caption{Income Distribution for the Three Groups in Rural Areas}
         \label{fig:1b}
     \end{subfigure}
     \caption{Income Distribution by Ethnic Group in Urban and Rural Areas}
    \label{fig:1}
\end{figure}

Table \ref{tab:1} is a summary of the descriptive statistics of the selected sample of the CHIP2018 data categorized by ethnicity and residential area. Among the 6,437 individuals being studied, 5,019 are Han, 772 are Mongols, and 646 are Manchus, which comprise approximately 78 percent, 12 percent, and 10 percent of the total sample respectively. Figure \ref{fig:1a} and Figure \ref{fig:1b} show the income distribution for the 3 ethnic groups in rural and urban regions respectively. Among these three ethnic groups, Mongols have the highest average individual annual income of ¥51,023 and Manchus have the lowest average individual annual income of ¥38,283. Mongols also have the largest years of schooling and family size, the lowest average age, and are most likely to have a job in the public sector or to be part of the Chinese Communist Party (CCP). Some of these characteristics and trends of Mongols can be explained by the affirmative actions within the autonomous region of Inner Mongolia. The government’s inclination towards hiring Mongols is presented in the higher proportion of Mongols in the public sector. In addition, a CCP membership is preferred for acquiring a government position in China, which explains the higher Communist percentage of Mongols. It can also be predicted that the difference in involvement in the public sector and the Communist party can account for a portion of the income gap.

In terms of residential areas, 60 percent of the total sample selected live in urban areas while 40 percent live in rural areas. Comparing rural and urban observations, the two regions do not differ much regarding gender proportions and age distributions. However, the average annual income in urban regions is almost twice the average annual income in rural regions. The urban-rural disparity is the largest for Mongols and the smallest for Manchus. With respect to other characteristics, the individuals from urban areas on average spend 3 more years in school and are more likely to be affiliated with the CCP and to hold a position in the government than rural individuals. Rural households also appear to have slightly larger family sizes than urban households.

\section{Methods}\label{sec4}

Three types of regression models are used to analyze the disparity in income between Han, Manchus, and Mongols in rural and urban regions separately. 

\subsection{Simple OLS Linear Regression}\label{subsec4.1}
The first method uses a simple OLS linear regression model with log income as the response variable and related endowment covariates as the independent variables. In R, the lm method is used to fit the following regression form:
\begin{gather*}
\label{eq:1}
    ln(wage_i) = \beta_0+\beta_1\times Age_i + \beta_2\times YearsofSchooling_i + \beta_3\times Ethnicity_i \\
    + \beta_4\times FamilySize_i
    + \beta_5\times Gender_i + \beta_6\times OccupationalSector_i \\
    + \beta_7\times PoliticalAffiliation_i +\epsilon_i
\end{gather*}
Here, 7 independent variables are included. $Age$, $YearsofSchooling$, and $FamilySize$ are all integer variables indicating the individual’s age, years of education received, and number of total family members respectively. $Ethnicity$ is a categorical variable with classes Han, Mongol, and Manchu. $Gender$ is a binary variable with 1 being male and 0 being female. For the binary variable $OccupationalSector$, I coded all the individuals employed in the public sector to be 1 and in the private sector to be 0 during the data preprocessing stage. Similarly, for the binary variable $PoliticalAffiliation$, I assigned a value of 0 to all the individuals who are not affiliated with the CCP and a value of 1 to individuals involved in the CCP. The response variable is the natural log of income of the individual for easier interpretation of the coefficient estimates. Two separate regressions are run for the rural and the urban samples.

\subsection{OLS Linear Regression with Interaction Terms}\label{subsec4.2}
Building on top of the first model, the second model uses an OLS linear regression that adds logically important interaction terms with ethnic status, taking the following forms:

\begin{gather*}
\label{eq:2}
    ln(wage_i) = \beta_0+\beta_1\times Age_i + \beta_2\times YearsofSchooling_i + \beta_3\times Ethnicity_i \\
    + \beta_4\times FamilySize_i
    + \beta_5\times Gender_i + \beta_6\times OccupationalSector_i \\
    + \beta_7\times PoliticalAffiliation_i +\epsilon_i
\end{gather*}

While the first 7 terms in this model remain the same as in the first linear regression model, 4 additional first-order interaction terms are added to further compare the returns to gender, education, occupation, and political affiliation for the three different ethnic groups. Five different models are used, with the first four each including a different interaction term and the last one including all interactions as the form specified in the equation above. 

\subsection{Blinder-Oaxaca Decomposition}\label{subsec4.3}
The third approach used in this study is the twofold Blinder-Oaxaca decomposition. As a useful tool in investigating the gap in the mean value differences of the response variable between two groups, the Blinder-Oaxaca method dissects the wage differential into two segments: one explained by the dependent variables included in the first OLS model and another attributable to discrimination. From running separate OLS regressions for each ethnic group specified in model 1, we can get estimates $b_{Han}$, $b_{Mongol}$ and $b_{Manchu}$ for coefficient vectors $\beta_{Han}$, $\beta_{Mongol}$, and $\beta_{Manchu}$. The form of the decomposition for the income gap between Han and Mongol is as follows, and the other pairwise comparisons are analogous:

\begin{gather*}
\label{eq:3}
\overline{ln⁡(wage_{Han})-ln⁡(wage_{Mongol})}\\
=b_{Han} \overline{X_{Han}}-b_{Mongol} \overline{X_{Mongol}}\\
=b_{Han}(\overline{X_{Han}}-\overline{X_{Mongol}})+\overline{X_{Mongol}}(b_{Han}-b_{Mongol})	
\end{gather*}

Here, the first half of Eq. \ref{eq:3} is the explained part of the income gap between Han and Mongols while the second half of Eq. \ref{eq:3} is the part accounting for the discriminatory factors. The R package oaxaca is used to carry out the decomposition analysis. Specifically, I used the Neumark Decomposition, which employs a pooled regression that does not include the ethnicity indicator as an independent variable, by specifying weights = -1 in the oaxaca function. Because the method can only take two groups, I conducted six Blinder-Oaxaca decomposition analyses to explore the wage gaps between the Han and the two minority groups separately and between the Mongols and the Manchus in urban and rural settings. 

\begin{figure}
     \centering
     \begin{subfigure}[b]{15cm}
         \centering
         \includegraphics[width=15cm]{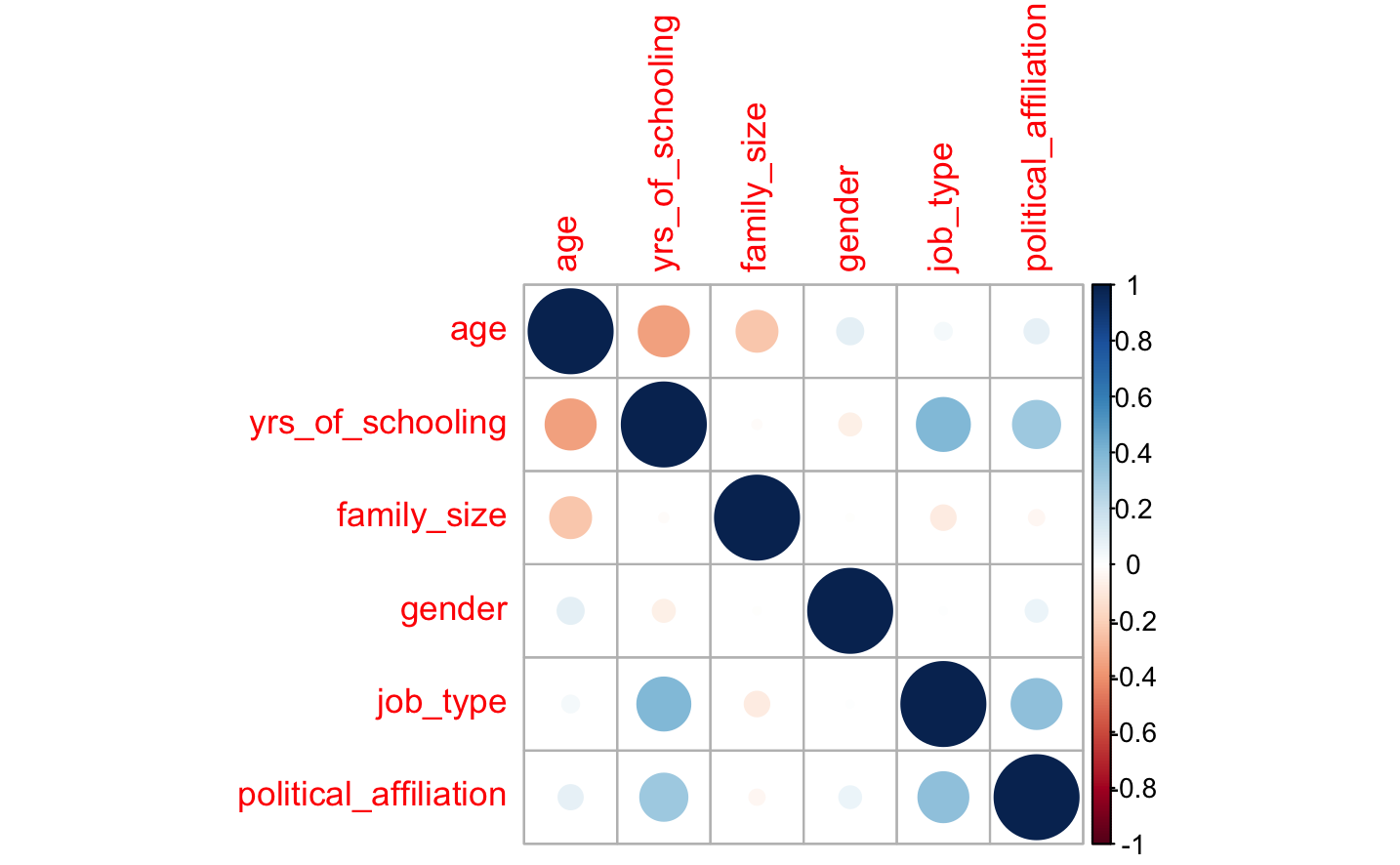}
         \caption{Correlation Plot for Urban Sample}
         \label{fig:2a}
     \end{subfigure}
     \vfill
     \begin{subfigure}[b]{15cm}
         \centering
         \includegraphics[width=15cm]{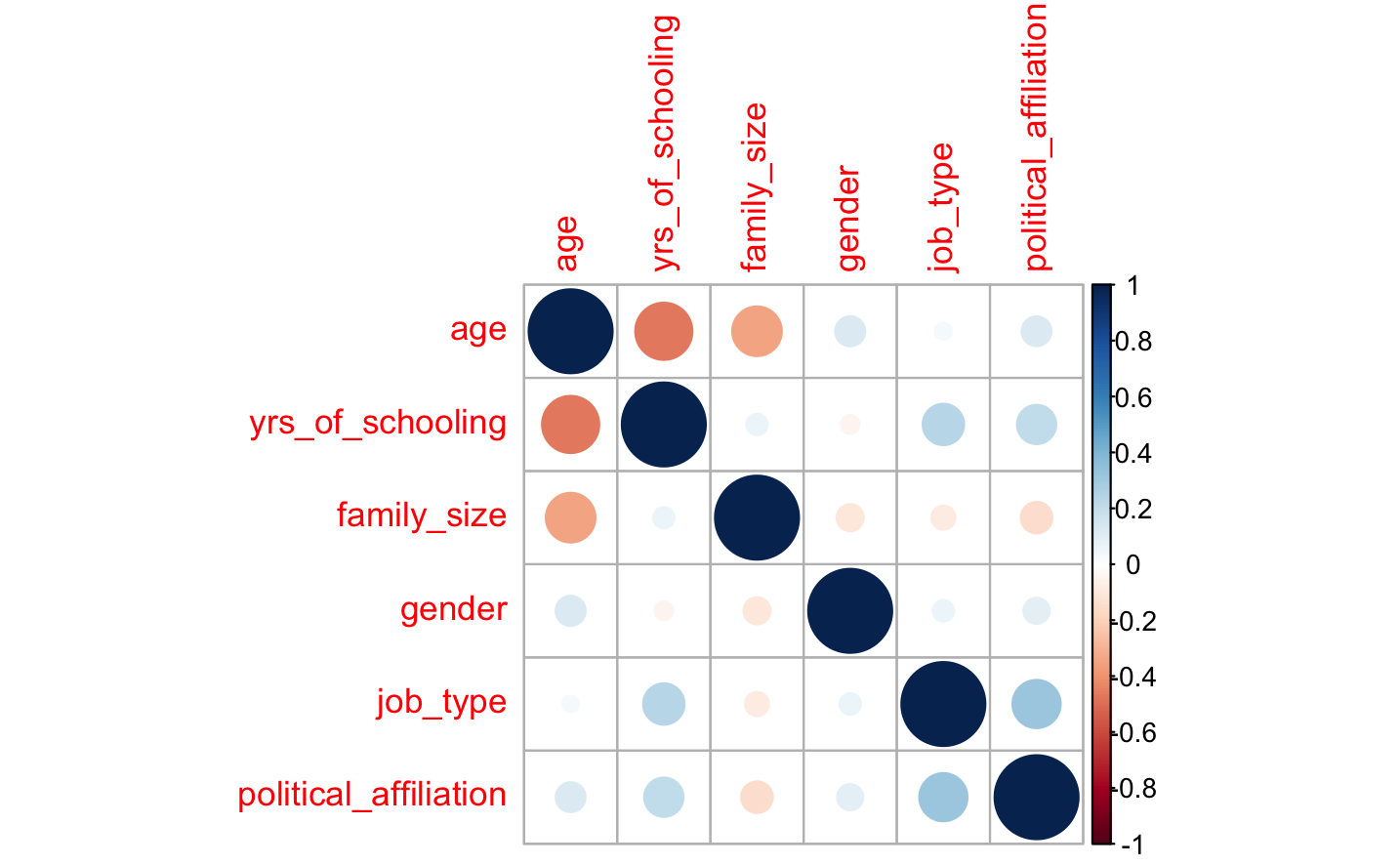}
         \caption{Correlation Plot for Rural Sample}
         \label{fig:2b}
     \end{subfigure}
    \caption{Correlation Plots}
    \label{fig:2}
\end{figure}

\section{Results}\label{sec5}
The findings from the three different analyses all suggest a significant effect of “outsider” ethnicity status on the income level of the rural population in Liaoning and Inner Mongolia.

\subsection{Simple OLS Linear Regression}\label{subsec5.1}

Figure \ref{fig:2a} is a correlation plot of all the included covariates in the OLS regression for the urban sample and Figure \ref{fig:2b} for the rural sample. At a cutoff value of 0.8, none of the variable pairs are too correlated to cause the problem of multicollinearity. The first columns of Table \ref{tab:2} and Table \ref{tab:3} present the results from the simple OLS linear regression model. In both the urban and rural regions, age, gender, and education serve as significant factors influencing the individual income level. Age and gender both negatively relate to income while education length positively affects income. The coefficient estimates in rural areas are larger in absolute value than the estimates in urban areas, indicating a larger income penalty for females, older, and less educated people in rural labor markets. Specifically, one more year in school increases the income by 4.2\% in urban areas and by 6.0\% in rural areas.

The models for rural and urban areas diverge in terms of the effects of ethnicity, job type, and political affiliation on an individual’s wage. In urban areas, the regression results indicate that holding a public sector job is associated with a 23.6\% increase in log income, which translates to approximately a 26.6\% higher actual income. Similarly, being a member of the Communist Party is linked to a 17\% increase in log income, corresponding to about an 18.5\% rise in actual income. These two covariates do not significantly affect the earnings levels of rural individuals. In rural areas, the Manchus do not differ from the Han regarding income after accounting for the other factors. However, the coefficient for Mongol status is -0.302, significant at a 0.05 level, meaning that the Mongols earned 26\% less than Hans after taking out the difference explained by the personal endowments. 

A stepwise backward elimination variable selection process is performed on the two linear models. This algorithm iteratively eliminates one variable at a time based on the model AIC values to choose the optimal model that included the most contributive variables that explain the most variation in the response variable. Based on the model selection results, for the urban model, the full model is the optimal model that minimizes AIC; for the rural model, the optimal reduced model takes the form:
\begin{gather*}
\ln⁡(wage_i)=\beta_0+\beta_1*Age_i+\beta_2*Education_i+\beta_3*Ethnicity_i+ \beta_4*Gender_i \\
    +\beta_5*PoliticalAffiliation_i+\epsilon
\end{gather*}
Compared to the full model, the reduced model gets rid of variables $FamilySize$ and $JobType$, both of which are proven to be statistically insignificant in the summary for the full model.

A possible explanation for this difference in significant coefficients between urban and rural areas is the implementation of affirmative action policies in the Inner Mongolia autonomous region. As discussed previously, there is a requirement for the number of Mongols employed in the Inner Mongolia government. Therefore, in urban areas, the higher income levels of Mongols can be explained by the preferential hiring decisions in the public sector. This conjecture is further examined in the next model, where interaction terms between ethnicity and job sector, and CCP affiliation are included. 

\begin{table}[p]
\centering
\captionsetup{aboveskip=4pt, belowskip=6pt}
\caption{OLS Linear Regression Results for Urban Area}
\label{tab:2}
\begin{threeparttable}
\setlength{\tabcolsep}{5pt}
\renewcommand{\arraystretch}{1.15}
\footnotesize
\begin{adjustbox}{width=0.92\textwidth,center}
\begin{tabular}{@{}lcccccc@{}}
\textbf{} \\
  \toprule &
  \multicolumn{6}{c}{\textbf{Urban}} \\
  \cmidrule(lr){2-7}
 &
  (1) &
  (2) &
  (3) &
  (4) &
  (5) &
  (6) \\
  \midrule
Age &
  \begin{tabular}[c]{@{}c@{}}-0.007**\\      (0.003)\end{tabular} &
  \begin{tabular}[c]{@{}c@{}}-0.007**\\      (0.003)\end{tabular} &
  \begin{tabular}[c]{@{}c@{}}-0.007**\\      (0.003)\end{tabular} &
  \begin{tabular}[c]{@{}c@{}}-0.007**\\      (0.003)\end{tabular} &
  \begin{tabular}[c]{@{}c@{}}-0.007**\\      (0.003)\end{tabular} &
  \begin{tabular}[c]{@{}c@{}}-0.007**\\      (0.003)\end{tabular} \\
Female &
  \begin{tabular}[c]{@{}c@{}}-0.380***\\      (0.049)\end{tabular} &
  \begin{tabular}[c]{@{}c@{}}-0.375***\\      (0.052)\end{tabular} &
  \begin{tabular}[c]{@{}c@{}}-0.380***\\      (0.048)\end{tabular} &
  \begin{tabular}[c]{@{}c@{}}-0.380***\\      (0.048)\end{tabular} &
  \begin{tabular}[c]{@{}c@{}}-0.380***\\      (0.048)\end{tabular} &
  \begin{tabular}[c]{@{}c@{}}-0.374***\\      (0.052)\end{tabular} \\
Mongol &
  \begin{tabular}[c]{@{}c@{}}0.131\\      (0.084)\end{tabular} &
  \begin{tabular}[c]{@{}c@{}}0.147\\      (0.117)\end{tabular} &
  \begin{tabular}[c]{@{}c@{}}0.127\\      (0.329)\end{tabular} &
  \begin{tabular}[c]{@{}c@{}}0.144\\      (0.120)\end{tabular} &
  \begin{tabular}[c]{@{}c@{}}0.150\\      (0.099)\end{tabular} &
  \begin{tabular}[c]{@{}c@{}}0.130\\      (0.340)\end{tabular} \\
Manchu &
  \begin{tabular}[c]{@{}c@{}}-0.132\\      (0.098)\end{tabular} &
  \begin{tabular}[c]{@{}c@{}}-0.116\\      (0.138)\end{tabular} &
  \begin{tabular}[c]{@{}c@{}}-0.172\\      (0.340)\end{tabular} &
  \begin{tabular}[c]{@{}c@{}}-0.173\\      (0.109)\end{tabular} &
  \begin{tabular}[c]{@{}c@{}}-0.135\\      (0.103)\end{tabular} &
  \begin{tabular}[c]{@{}c@{}}-0.113\\      (0.351)\end{tabular} \\
Years of Schooling &
  \begin{tabular}[c]{@{}c@{}}0.042***\\      (0.008)\end{tabular} &
  \begin{tabular}[c]{@{}c@{}}0.043***\\      (0.008)\end{tabular} &
  \begin{tabular}[c]{@{}c@{}}0.042***\\      (0.009)\end{tabular} &
  \begin{tabular}[c]{@{}c@{}}0.042***\\      (0.008)\end{tabular} &
  \begin{tabular}[c]{@{}c@{}}0.042***\\      (0.008)\end{tabular} &
  \begin{tabular}[c]{@{}c@{}}0.042***\\      (0.009)\end{tabular} \\
Family Size &
  \begin{tabular}[c]{@{}c@{}}0.053\\      (0.029)\end{tabular} &
  \begin{tabular}[c]{@{}c@{}}0.052\\      (0.029)\end{tabular} &
  \begin{tabular}[c]{@{}c@{}}0.053\\      (0.029)\end{tabular} &
  \begin{tabular}[c]{@{}c@{}}0.053\\      (0.029)\end{tabular} &
  \begin{tabular}[c]{@{}c@{}}0.053\\      (0.029)\end{tabular} &
  \begin{tabular}[c]{@{}c@{}}0.053\\      (0.029)\end{tabular} \\
Public/Gov Job &
  \begin{tabular}[c]{@{}c@{}}0.236***\\      (0.058)\end{tabular} &
  \begin{tabular}[c]{@{}c@{}}0.236***\\      (0.058)\end{tabular} &
  \begin{tabular}[c]{@{}c@{}}0.236***\\      (0.058)\end{tabular} &
  \begin{tabular}[c]{@{}c@{}}0.228***\\      (0.062)\end{tabular} &
  \begin{tabular}[c]{@{}c@{}}0.236***\\      (0.058)\end{tabular} &
  \begin{tabular}[c]{@{}c@{}}0.226***\\      (0.063)\end{tabular} \\
Communist &
  \begin{tabular}[c]{@{}c@{}}0.170*\\      (0.069)\end{tabular} &
  \begin{tabular}[c]{@{}c@{}}0.169*\\      (0.069)\end{tabular} &
  \begin{tabular}[c]{@{}c@{}}0.170*\\      (0.069)\end{tabular} &
  \begin{tabular}[c]{@{}c@{}}0.172*\\      (0.069)\end{tabular} &
  \begin{tabular}[c]{@{}c@{}}0.177*\\      (0.075)\end{tabular} &
  \begin{tabular}[c]{@{}c@{}}0.183*\\      (0.076)\end{tabular} \\
 &
   &
   &
   &
   &
   &
   \\
Mongol * Female &
   &
  \begin{tabular}[c]{@{}c@{}}-0.034\\      (0.166)\end{tabular} &
   &
   &
   &
  \begin{tabular}[c]{@{}c@{}}-0.042\\      (0.169)\end{tabular} \\
Manchu * Female &
   &
  \begin{tabular}[c]{@{}c@{}}-0.032\\      (0.195)\end{tabular} &
   &
   &
   &
  \begin{tabular}[c]{@{}c@{}}-0.031\\      (0.201)\end{tabular} \\
 &
   &
   &
   &
   &
   &
   \\
Mongol *   Years of Schooling &
   &
   &
  \begin{tabular}[c]{@{}c@{}}0.000\\      (0.025)\end{tabular} &
   &
   &
  \begin{tabular}[c]{@{}c@{}}0.004\\      (0.029)\end{tabular} \\
Manchu *   Years of Schooling &
   &
   &
  \begin{tabular}[c]{@{}c@{}}0.004\\      (0.028)\end{tabular} &
   &
   &
  \begin{tabular}[c]{@{}c@{}}-0.004\\      (0.032)\end{tabular} \\
 &
   &
   &
   &
   &
   &
   \\
Mongol *   Public/Gov Job &
   &
   &
   &
  \begin{tabular}[c]{@{}c@{}}-0.022\\      (0.167)\end{tabular} &
   &
  \begin{tabular}[c]{@{}c@{}}-0.007\\      (0.199)\end{tabular} \\
Manchu *   Public/Gov Job &
   &
   &
   &
  \begin{tabular}[c]{@{}c@{}}0.209\\      (0.244)\end{tabular} &
   &
  \begin{tabular}[c]{@{}c@{}}0.225\\      (0.261)\end{tabular} \\
 &
   &
   &
   &
   &
   &
   \\
Mongol *   Communist &
   &
   &
   &
   &
  \begin{tabular}[c]{@{}c@{}}-0.068\\      (0.185)\end{tabular} &
  \begin{tabular}[c]{@{}c@{}}-0.075\\      (0.201)\end{tabular} \\
Manchu *   Communist &
   &
   &
   &
   &
  \begin{tabular}[c]{@{}c@{}}0.033\\      (0.314)\end{tabular} &
  \begin{tabular}[c]{@{}c@{}}0.007\\      (0.329)\end{tabular} \\
 &
   &
   &
   &
   &
   &
   \\
Observations &
  1941 &
  1941 &
  1941 &
  1941 &
  1941 &
  1941 \\
R-squared &
  0.094 &
  0.094 &
  0.094 &
  0.095 &
  0.094 &
  0.095 \\
Adjusted R-squared &
  0.091 &
  0.090 &
  0.090 &
  0.090 &
  0.090 &
  0.087 \\
  \bottomrule
\end{tabular}
\end{adjustbox}

\begin{tablenotes}[flushleft]\footnotesize
\item Significance codes: *** $p<0.001$, ** $p<0.01$, * $p<0.05$.
\item Standard errors in parentheses.
\end{tablenotes}
\end{threeparttable}
\end{table}

\begin{table}[p]
\centering
\captionsetup{aboveskip=4pt, belowskip=6pt}
\caption{OLS Linear Regression Results for Rural Area}
\label{tab:3}
\begin{threeparttable}
\setlength{\tabcolsep}{5pt}
\renewcommand{\arraystretch}{1.15}
\footnotesize
\begin{adjustbox}{width=0.92\textwidth,center}
\begin{tabular}{@{}lcccccc@{}}
\textbf{}
\\
\toprule  &
  \multicolumn{6}{c}{\textbf{Rural}} \\
  \cmidrule(lr){2-7}
 &
  (1) &
  (2) &
  (3) &
  (4) &
  (5) &
  (6) \\
  \midrule
Age &
  \begin{tabular}[c]{@{}c@{}}-0.012**\\ (0.004)\end{tabular} &
  \begin{tabular}[c]{@{}c@{}}-0.012**\\ (0.004)\end{tabular} &
  \begin{tabular}[c]{@{}c@{}}-0.013**\\ (0.004)\end{tabular} &
  \begin{tabular}[c]{@{}c@{}}-0.013**\\ (0.004)\end{tabular} &
  \begin{tabular}[c]{@{}c@{}}-0.013**\\ (0.004)\end{tabular} &
  \begin{tabular}[c]{@{}c@{}}-0.013**\\ (0.004)\end{tabular} \\
Female &
  \begin{tabular}[c]{@{}c@{}}-0.542***\\ (0.091)\end{tabular} &
  \begin{tabular}[c]{@{}c@{}}-0.572***\\ (0.106)\end{tabular} &
  \begin{tabular}[c]{@{}c@{}}-0.541***\\ (0.091)\end{tabular} &
  \begin{tabular}[c]{@{}c@{}}-0.532***\\ (0.091)\end{tabular} &
  \begin{tabular}[c]{@{}c@{}}-0.545***\\ (0.091)\end{tabular} &
  \begin{tabular}[c]{@{}c@{}}-0.576***\\ (0.106)\end{tabular} \\
Mongol &
  \begin{tabular}[c]{@{}c@{}}-0.302*\\ (0.141)\end{tabular} &
  \begin{tabular}[c]{@{}c@{}}-0.309.\\ (0.173)\end{tabular} &
  \begin{tabular}[c]{@{}c@{}}-0.612\\ (0.375)\end{tabular} &
  \begin{tabular}[c]{@{}c@{}}-0.441**\\ (0.162)\end{tabular} &
  \begin{tabular}[c]{@{}c@{}}-0.427**\\ (0.153)\end{tabular} &
  \begin{tabular}[c]{@{}c@{}}-0.452\\ (0.404)\end{tabular} \\
Manchu &
  \begin{tabular}[c]{@{}c@{}}0.106\\ (0.110)\end{tabular} &
  \begin{tabular}[c]{@{}c@{}}0.060\\ (0.130)\end{tabular} &
  \begin{tabular}[c]{@{}c@{}}-0.103\\ (0.378)\end{tabular} &
  \begin{tabular}[c]{@{}c@{}}-0.101\\ (0.117)\end{tabular} &
  \begin{tabular}[c]{@{}c@{}}0.122\\ (0.113)\end{tabular} &
  \begin{tabular}[c]{@{}c@{}}-0.139\\ (0.389)\end{tabular} \\
Years of Schooling &
  \begin{tabular}[c]{@{}c@{}}0.060***\\ (0.017)\end{tabular} &
  \begin{tabular}[c]{@{}c@{}}0.060***\\ (0.017)\end{tabular} &
  \begin{tabular}[c]{@{}c@{}}0.051**\\ (0.019)\end{tabular} &
  \begin{tabular}[c]{@{}c@{}}0.057***\\ (0.017)\end{tabular} &
  \begin{tabular}[c]{@{}c@{}}0.055**\\ (0.017)\end{tabular} &
  \begin{tabular}[c]{@{}c@{}}0.052**\\ (0.019)\end{tabular} \\
Family Size &
  \begin{tabular}[c]{@{}c@{}}-0.009\\ (0.038)\end{tabular} &
  \begin{tabular}[c]{@{}c@{}}-0.010\\ (0.038)\end{tabular} &
  \begin{tabular}[c]{@{}c@{}}-0.008\\ (0.038)\end{tabular} &
  \begin{tabular}[c]{@{}c@{}}-0.012\\ (0.038)\end{tabular} &
  \begin{tabular}[c]{@{}c@{}}-0.008\\ (0.038)\end{tabular} &
  \begin{tabular}[c]{@{}c@{}}-0.011\\ (0.038)\end{tabular} \\
Public/Gov Job &
  \begin{tabular}[c]{@{}c@{}}0.093\\ (0.132)\end{tabular} &
  \begin{tabular}[c]{@{}c@{}}0.094\\ (0.133)\end{tabular} &
  \begin{tabular}[c]{@{}c@{}}0.090\\ (0.132)\end{tabular} &
  \begin{tabular}[c]{@{}c@{}}-0.002\\ (0.157)\end{tabular} &
  \begin{tabular}[c]{@{}c@{}}0.069\\ (0.132)\end{tabular} &
  \begin{tabular}[c]{@{}c@{}}0.016\\ (0.160)\end{tabular} \\
Communist &
  \begin{tabular}[c]{@{}c@{}}0.198\\ (0.156)\end{tabular} &
  \multicolumn{1}{l}{\begin{tabular}[c]{@{}l@{}}0.199\\ (0.156)\end{tabular}} &
  \multicolumn{1}{l}{\begin{tabular}[c]{@{}l@{}}0.187\\ (0.157)\end{tabular}} &
  \multicolumn{1}{l}{\begin{tabular}[c]{@{}l@{}}0.181\\ (0.157)\end{tabular}} &
  \begin{tabular}[c]{@{}c@{}}0.123\\ (0.178)\end{tabular} &
  \begin{tabular}[c]{@{}c@{}}0.142\\ (0.182)\end{tabular} \\
\multicolumn{1}{c}{} &
  \multicolumn{1}{l}{} &
  \multicolumn{1}{l}{} &
   &
   &
  \multicolumn{1}{l}{} &
  \multicolumn{1}{l}{} \\
Mongol * Female &
   &
  \begin{tabular}[c]{@{}c@{}}0.022\\ (0.298)\end{tabular} &
   &
   &
   &
  \begin{tabular}[c]{@{}c@{}}0.155\\ (0.311)\end{tabular} \\
Manchu * Female &
   &
  \begin{tabular}[c]{@{}c@{}}0.165\\ (0.244)\end{tabular} &
   &
   &
   &
  \begin{tabular}[c]{@{}c@{}}0.137\\ (0.247)\end{tabular} \\
 &
   &
   &
   &
   &
   &
   \\
Mongol * Years of Schooling &
   &
   &
  \begin{tabular}[c]{@{}c@{}}0.033\\ (0.037)\end{tabular} &
   &
   &
  \begin{tabular}[c]{@{}c@{}}-0.009\\ (0.042)\end{tabular} \\
Manchu * Years of Schooling &
   &
   &
  \begin{tabular}[c]{@{}c@{}}0.024\\ (0.042)\end{tabular} &
   &
   &
  \begin{tabular}[c]{@{}c@{}}0.026\\ (0.044)\end{tabular} \\
 &
   &
   &
   &
   &
   &
   \\
Mongol * Public/Gov Job &
   &
   &
   &
  \begin{tabular}[c]{@{}c@{}}0.580.\\ (0.335)\end{tabular} &
   &
  \begin{tabular}[c]{@{}c@{}}0.355\\ (0.403)\end{tabular} \\
Manchu * Public/Gov Job &
   &
   &
   &
  \begin{tabular}[c]{@{}c@{}}0.052\\ (0.333)\end{tabular} &
   &
  \begin{tabular}[c]{@{}c@{}}0.026\\ (0.350)\end{tabular} \\
 &
   &
   &
   &
   &
   &
   \\
Mongol * Communist &
   &
   &
   &
   &
  \begin{tabular}[c]{@{}c@{}}0.767*\\ (0.381)\end{tabular} &
  \begin{tabular}[c]{@{}c@{}}0.641\\ (0.482)\end{tabular} \\
Manchu * Communist &
   &
   &
   &
   &
  \begin{tabular}[c]{@{}c@{}}-0.337\\ (0.461)\end{tabular} &
  \begin{tabular}[c]{@{}c@{}}-0.335\\ (0.472)\end{tabular} \\
 &
   &
   &
   &
   &
   &
   \\
Observations &
  724 &
  724 &
  724 &
  724 &
  724 &
  724 \\
R-squared &
  0.110 &
  0.110 &
  0.111 &
  0.113 &
  0.116 &
  0.118 \\
Adjusted R-squared &
  0.100 &
  0.098 &
  0.098 &
  0.101 &
  0.104 &
  0.098\\
    \bottomrule
\end{tabular}
\end{adjustbox}

\begin{tablenotes}[flushleft]\footnotesize
\item Significance codes: *** $p<0.001$, ** $p<0.01$, * $p<0.05$.
\item Standard errors in parentheses.
\end{tablenotes}
\end{threeparttable}
\end{table}

\subsection{OLS Linear Regression with Interaction Terms}\label{subsec5.2}

The second to the last columns of Table \ref{tab:2} and Table \ref{tab:3} display the coefficient estimates from five models with additional interaction terms to the simple model \ref{eq:1}. In model 2, only the interaction between ethnicity and gender is added; in model 3, only the interaction between ethnicity and education length is added; in model 4, only the interaction between ethnicity and job sector is added; in model 5, only the interaction between ethnicity and political affiliation is added; in model 6, all previous interaction terms are included. In all models for the urban sample, none of the interaction terms significantly influence the income level. In models 2, 3, and 6 for the rural sample, the coefficients for the interaction terms are also not significant. 

In model 4 for the rural population, while the coefficient estimates for terms associated with Manchu status remain insignificant, the coefficient estimate for the interaction term between Mongol status and the position in a public/government job is 0.580, significant at 0.1 level. The coefficient estimate for the Mongol indicator variable is -0.441 with a 0.01 level of significance. This suggests that while Mongols who work in the private sector suffer from 35.7\% lower income compared to the other two groups, employment in the public sector increases Mongols’ income by 78.6\%, implying that Mongols in the government or public sector earn 15\% more than their Han counterparts. Similarly, in model 5 for the rural population, the coefficient estimate for Mongol status is -0.427 with 0.01 significance level and for the interaction term between Mongol status and the association with the CCP is 0.767, significant at 0.05 level. This reveals that Communist Mongols earn 40\% more than Communist Hans and 115\% more than Mongols who are not CCP members. 

The stepwise model selection algorithm is again applied on both the model with all interaction terms on the urban and the rural samples. Based on the variable selection results, the algorithm chooses model 1 (the model without any interaction terms) for the urban sample, which corresponds to the statistical insignificance for the coefficients on all the interaction terms. The algorithm selects model 5 (the model with interaction term between ethnicity and political affiliation) for the rural sample, which is confirmed by the highest statistical significance of the coefficient of interaction between Mongol status and political affiliation among all the coefficients for interaction terms.

Mongol’s higher return to public/government career further confirms the effectiveness of the preferential hiring policy established in the Regional Ethnic Autonomy Laws of China regarding alleviating the wage disparity caused by ethnic discrimination. The results demonstrate that by encouraging public firms to hire more Mongols, the approach increases the average Mongol earnings levels, consequently reducing the ethnic income gap.

\begin{figure}
     \centering
     \begin{subfigure}[b]{14cm}
         \centering
         \includegraphics[width=14cm]{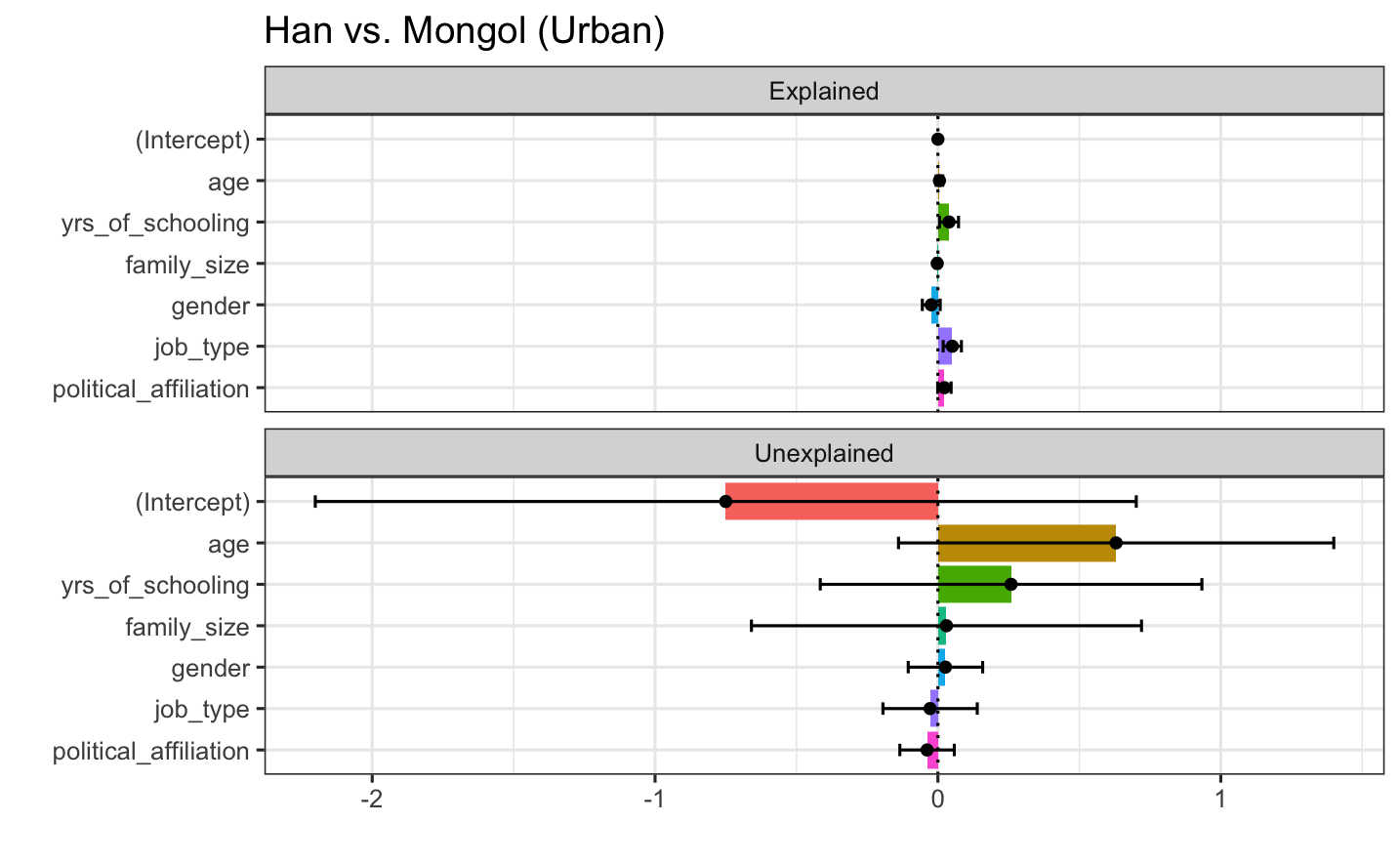}
         \caption{Blinder-Oaxaca Decomposition Graph for Han-Mongol Income Gaps in Urban Areas}
         \label{fig:3a}
     \end{subfigure}
     \vfill
     \begin{subfigure}[b]{14cm}
         \centering
         \includegraphics[width=14cm]{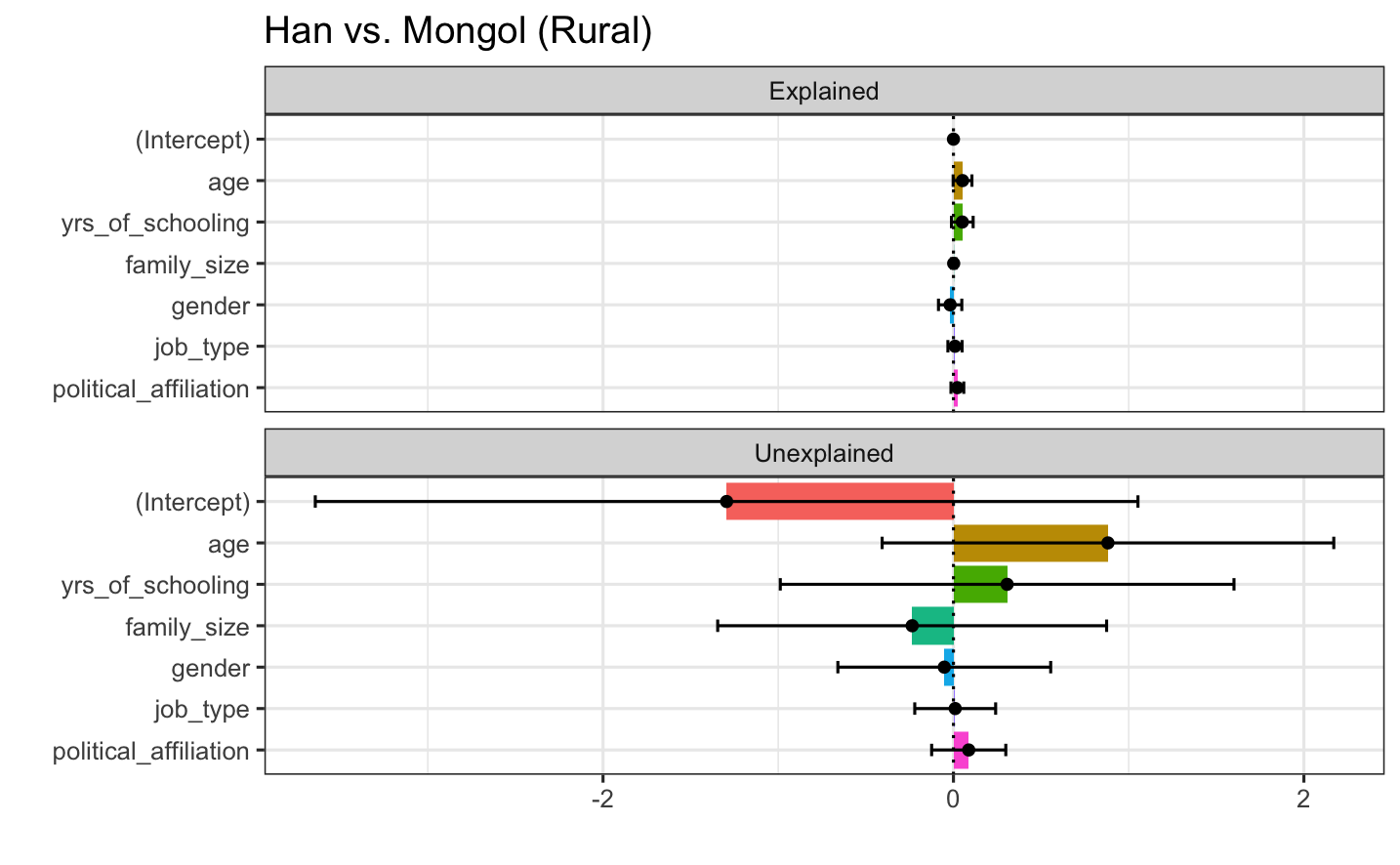}
         \caption{Blinder-Oaxaca Decomposition Graph for Han-Mongol Income Gaps in Rural Areas}
         \label{fig:3b}
     \end{subfigure}
    \caption{Blinder-Oaxaca Decomposition Graph for Han-Mongol Income Gaps in Urban and Rural Areas}
    \label{fig:3}
\end{figure}

\subsection{Blinder-Oaxaca Decomposition}\label{subsec5.3}
Table \ref{tab:4} and Table \ref{tab:5} presents the results from the six pairwise Blinder-Oaxaca decomposition analyses among the three ethnic groups in urban and rural regions separately. 

\begin{figure}
     \centering
     \begin{subfigure}[b]{14cm}
         \centering
         \includegraphics[width=14cm]{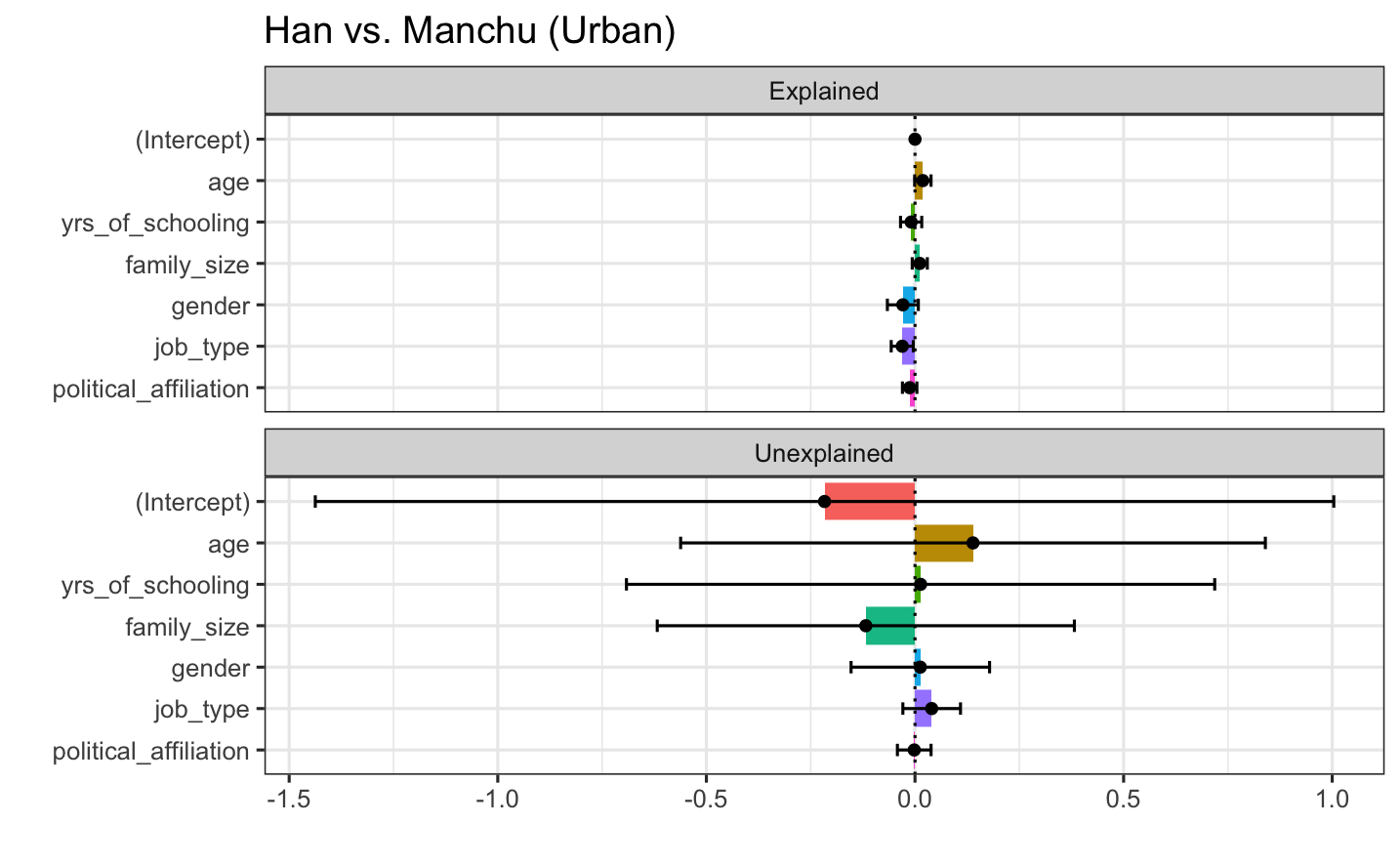}
         \caption{Blinder-Oaxaca Decomposition Graph for Han-Manchu Income Gaps in Urban Areas}
         \label{fig:4a}
     \end{subfigure}
     \vfill
     \begin{subfigure}[b]{14cm}
         \centering
         \includegraphics[width=14cm]{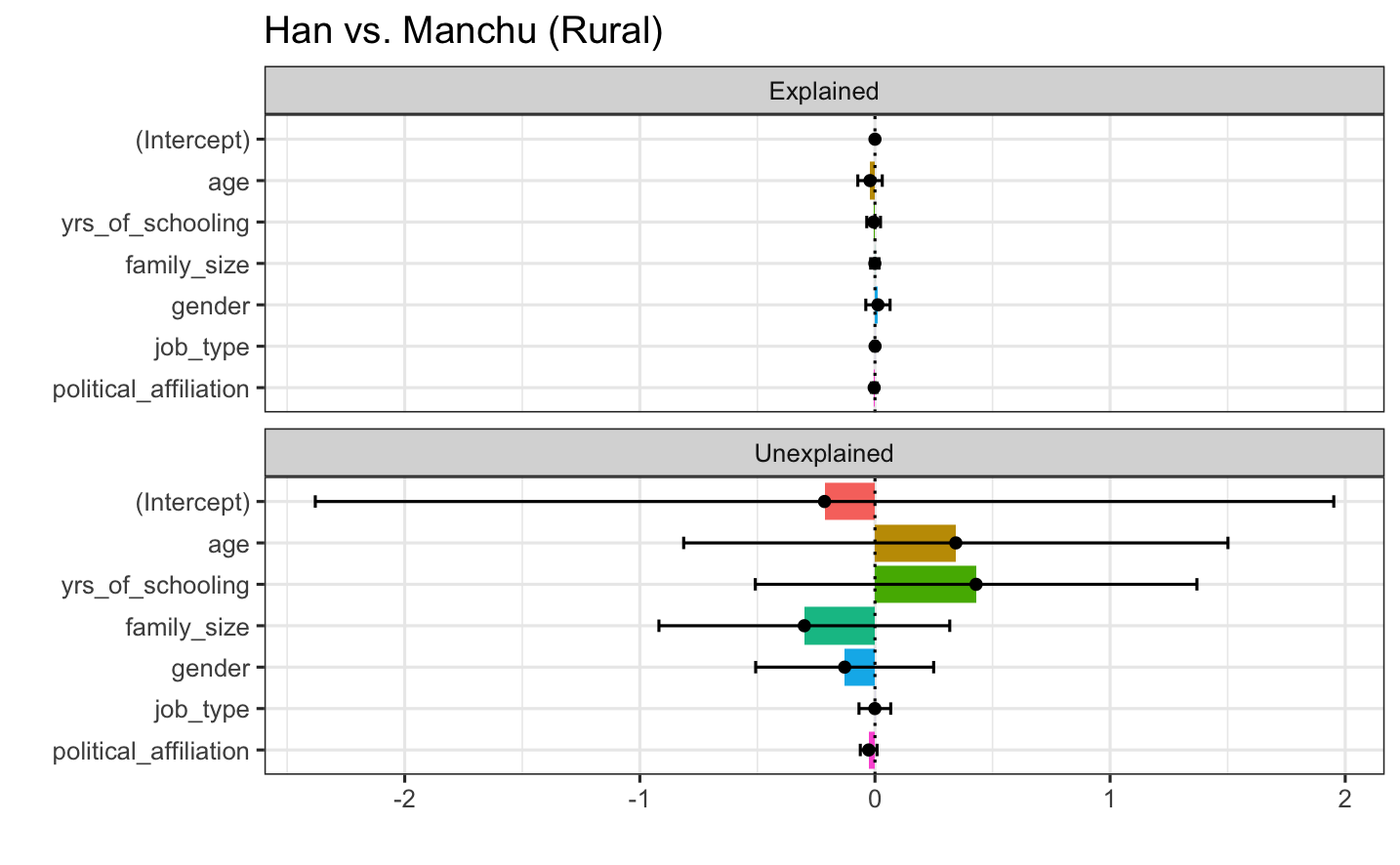}
         \caption{Blinder-Oaxaca Decomposition Graph for Han-Manchu Income Gaps in Rural Areas}
         \label{fig:4b}
     \end{subfigure}
    \caption{Blinder-Oaxaca Decomposition Graph for Han-Manchu Income Gaps in Urban and Rural Areas}
    \label{fig:4}
\end{figure}

\begin{table}[p]
\centering
\captionsetup{aboveskip=4pt, belowskip=6pt}
\caption{Blinder-Oaxaca Decomposition Results for Urban Area}
\label{tab:4}
\begin{threeparttable}
\setlength{\tabcolsep}{5pt}
\renewcommand{\arraystretch}{1.15}
\footnotesize
\begin{adjustbox}{width=\textwidth,center}
\begin{tabular}{@{}lcccccc@{}}
\\
  \toprule
\textbf{} &
  \textbf{} &
  \multicolumn{3}{c}{\textbf{Urban}} \\
 \cmidrule(lr){3-5}
 &
   &
  Han vs. Mongol &
  Han vs. Manchu &
  Manchu vs. Mongol \\
  \midrule
\multirow{2}{*}{Age} &
  Explained &
  \begin{tabular}[c]{@{}c@{}}0.005\\ (0.006)\end{tabular} &
  \begin{tabular}[c]{@{}c@{}}0.019\\ (0.011)\end{tabular} &
  \begin{tabular}[c]{@{}c@{}}0.004\\ (0.009)\end{tabular} \\
 &
  Unexplained &
  \begin{tabular}[c]{@{}c@{}}0.630\\ (0.372)\end{tabular} &
  \begin{tabular}[c]{@{}c@{}}0.139\\ (0.364)\end{tabular} &
  \begin{tabular}[c]{@{}c@{}}0.474\\ (0.469)\end{tabular} \\
\multirow{2}{*}{Gender} &
  Explained &
  \begin{tabular}[c]{@{}c@{}}-0.024\\ (0.014)\end{tabular} &
  \begin{tabular}[c]{@{}c@{}}-0.029\\ (0.016)\end{tabular} &
  \begin{tabular}[c]{@{}c@{}}0.006\\ (0.027)\end{tabular} \\
 &
  Unexplained &
  \begin{tabular}[c]{@{}c@{}}0.027\\ (0.075)\end{tabular} &
  \begin{tabular}[c]{@{}c@{}}0.012\\ (0.080)\end{tabular} &
  \begin{tabular}[c]{@{}c@{}}0.014\\ (0.102)\end{tabular} \\
\multirow{2}{*}{Years of Schooling} &
  Explained &
  \begin{tabular}[c]{@{}c@{}}0.039**\\ (0.014)\end{tabular} &
  \begin{tabular}[c]{@{}c@{}}-0.009\\ (0.013)\end{tabular} &
  \begin{tabular}[c]{@{}c@{}}0.066*\\ (0.028)\end{tabular} \\
 &
  Unexplained &
  \begin{tabular}[c]{@{}c@{}}0.259\\ (0.338)\end{tabular} &
  \begin{tabular}[c]{@{}c@{}}0.013\\ (0.341)\end{tabular} &
  \begin{tabular}[c]{@{}c@{}}0.228\\ (0.475)\end{tabular} \\
\multirow{2}{*}{Family Size} &
  Explained &
  \begin{tabular}[c]{@{}c@{}}-0.002\\ (0.003)\end{tabular} &
  \begin{tabular}[c]{@{}c@{}}0.011\\ (0.007)\end{tabular} &
  \begin{tabular}[c]{@{}c@{}}-0.006\\ (0.019)\end{tabular} \\
 &
  Unexplained &
  \begin{tabular}[c]{@{}c@{}}0.030\\ (0.350)\end{tabular} &
  \begin{tabular}[c]{@{}c@{}}-0.118\\ (0.251)\end{tabular} &
  \begin{tabular}[c]{@{}c@{}}0.141\\ (0.338)\end{tabular} \\
\multirow{2}{*}{Public/Gov Job} &
  Explained &
  \begin{tabular}[c]{@{}c@{}}0.051***\\ (0.015)\end{tabular} &
  \begin{tabular}[c]{@{}c@{}}-0.031**\\ (0.012)\end{tabular} &
  \begin{tabular}[c]{@{}c@{}}0.109*\\ (0.048)\end{tabular} \\
 &
  Unexplained &
  \begin{tabular}[c]{@{}c@{}}-0.028\\ (0.093)\end{tabular} &
  \begin{tabular}[c]{@{}c@{}}0.040\\ (0.034)\end{tabular} &
  \begin{tabular}[c]{@{}c@{}}-0.095\\ (0.083)\end{tabular} \\
\multirow{2}{*}{Communist} &
  Explained &
  \begin{tabular}[c]{@{}c@{}}0.023*\\ (0.011)\end{tabular} &
  \begin{tabular}[c]{@{}c@{}}-0.013*\\ (0.008)\end{tabular} &
  \begin{tabular}[c]{@{}c@{}}0.022\\ (0.029)\end{tabular} \\
 &
  Unexplained &
  \begin{tabular}[c]{@{}c@{}}-0.038\\ (0.052)\end{tabular} &
  \begin{tabular}[c]{@{}c@{}}-0.002\\ (0.020)\end{tabular} &
  \begin{tabular}[c]{@{}c@{}}-0.023\\ (0.035)\end{tabular} \\
 &
   &
   &
   &
   \\
\multirow{3}{*}{Average Log Income Differences} &
  Total &
  0.222 &
  -0.183 &
  0.405 \\
 &
  Explained &
  \begin{tabular}[c]{@{}c@{}}0.092***\\ (0.027)\end{tabular} &
  \begin{tabular}[c]{@{}c@{}}-0.051\\ (0.029)\end{tabular} &
  \begin{tabular}[c]{@{}c@{}}0.200***\\ (0.048)\end{tabular} \\
 &
  Unexplained &
  \begin{tabular}[c]{@{}c@{}}0.130*\\ (0.065)\end{tabular} &
  \begin{tabular}[c]{@{}c@{}}-0.131\\ (0.078)\end{tabular} &
  \begin{tabular}[c]{@{}c@{}}0.205*\\ (0.092)\end{tabular} \\
 &
   &
   &
   &
   \\
\multirow{3}{*}{Average Income Differences} &
  Total &
  9368.77* &
  -10973.46* &
  20343.23* \\
 &
  Explained &
  \begin{tabular}[c]{@{}c@{}}3077.66*\\ (1314.74)\end{tabular} &
  \begin{tabular}[c]{@{}c@{}}-2095.70.\\ (1151.58)\end{tabular} &
  \begin{tabular}[c]{@{}c@{}}6049.52*\\ (2590.38)\end{tabular} \\
 &
  Unexplained &
  \begin{tabular}[c]{@{}c@{}}6291.11*\\ (3140.87)\end{tabular} &
  \begin{tabular}[c]{@{}c@{}}-8877.76**\\ (2723.30)\end{tabular} &
  \begin{tabular}[c]{@{}c@{}}14292.71**\\ (4874.97)\end{tabular}\\
  \bottomrule
\end{tabular}
\end{adjustbox}

\begin{tablenotes}[flushleft]\footnotesize
\item Significance codes: *** $p<0.001$, ** $p<0.01$, * $p<0.05$.
\item Standard errors in parentheses.
\item In group A vs B, results are showing the gap = B - A.
\end{tablenotes}
\end{threeparttable}
\end{table}

\begin{figure}
     \centering
     \begin{subfigure}[b]{14cm}
         \centering
         \includegraphics[width=14cm]{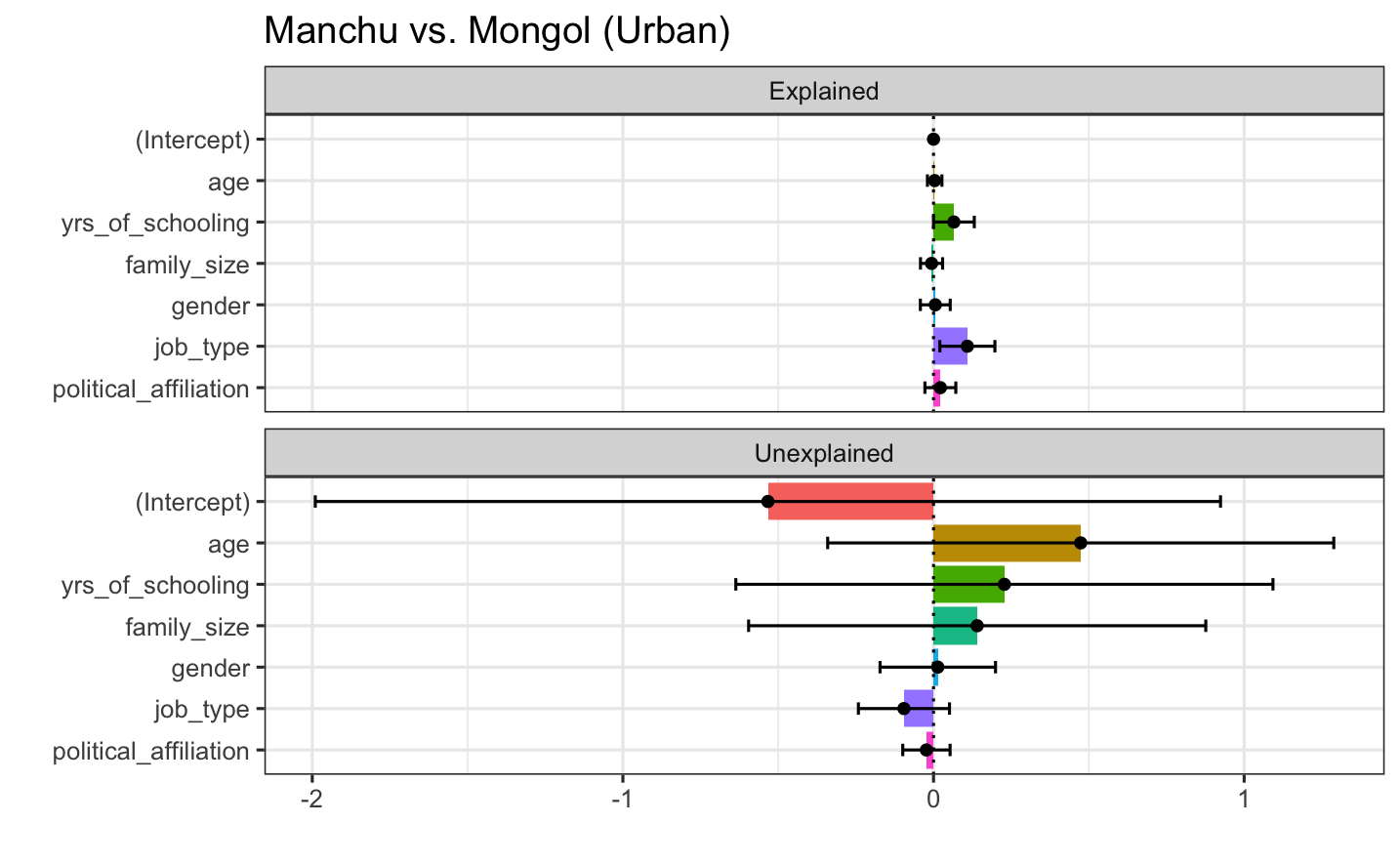}
         \caption{Blinder-Oaxaca Decomposition Graph for Manchu-Mongol Income Gaps in Urban Areas}
         \label{fig:5a}
     \end{subfigure}
     \vfill
     \begin{subfigure}[b]{14cm}
         \centering
         \includegraphics[width=14cm]{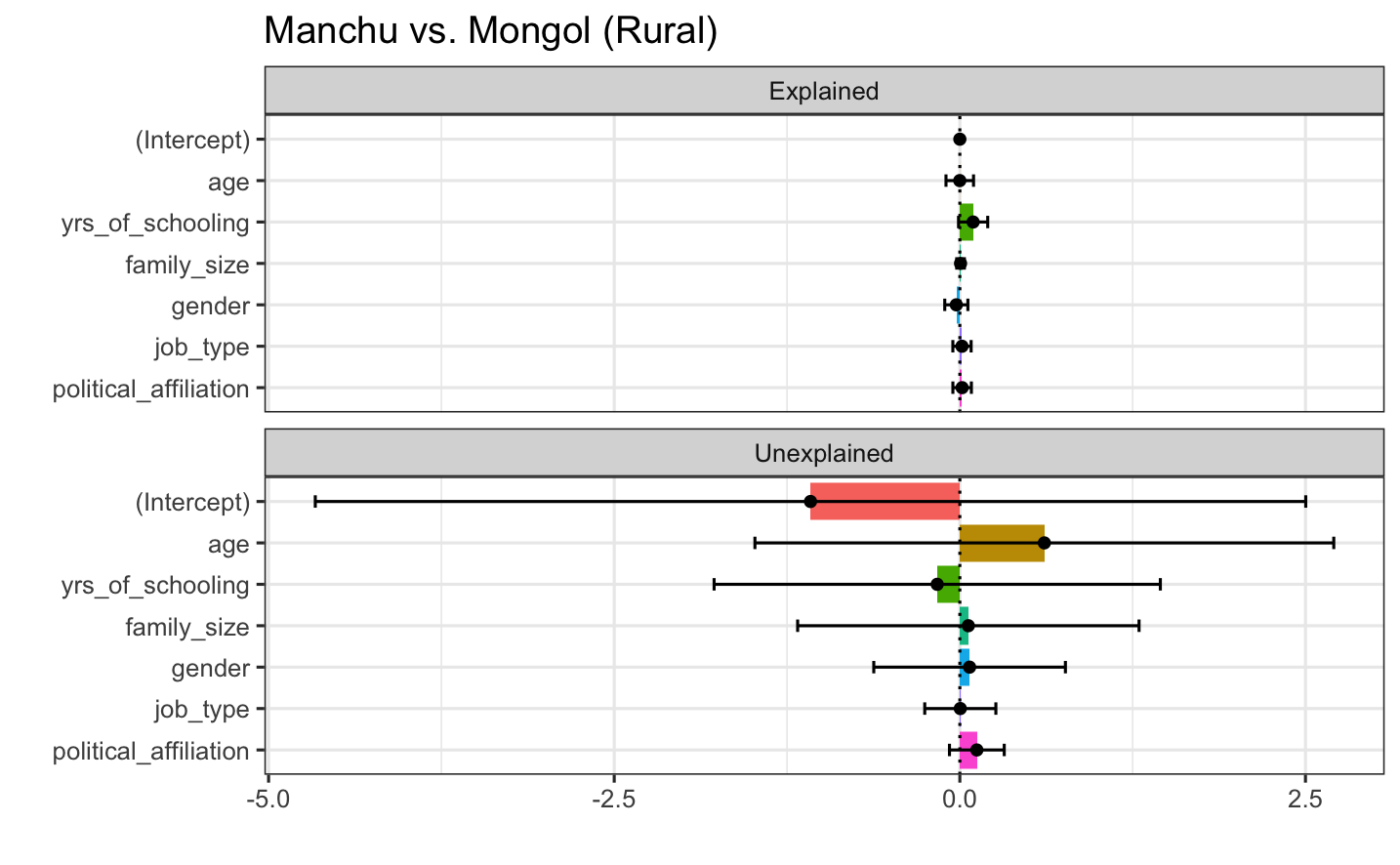}
         \caption{Blinder-Oaxaca Decomposition Graph for Manchu-Mongol Income Gaps in Rural Areas}
         \label{fig:5b}
     \end{subfigure}
    \caption{Blinder-Oaxaca Decomposition Graph for Manchu-Mongol Income Gaps in Urban and Rural Areas}
    \label{fig:5}
\end{figure}

\subsubsection{Han-Mongol Gap}\label{subsubsec1}
In urban areas, Mongols on average earn ¥9,368.77 more than Hans, where ¥3,077.66 of the gap is explained by the group differences in endowment and the rest ¥6,291.11 is attributed to possible discriminatory factors. Figure \ref{fig:3a} and \ref{fig:3b} plots the decomposition of the regression coefficients for each variable. As shown in the plot, much of the urban Han-Mongol earnings gap can be explained by the Mongols’ longer years of education, higher proportion in governmental or public sector positions, and higher probability of being a member of the Communist Party. 

The Han-Mongol wage gap is ¥1,362.01 in rural regions, slightly smaller compared to the urban regions. This gap can be decomposed into ¥2,376.70 as the result of the group differences in explanatory variables and -¥1,014.70 remains unexplained, although the unexplained part is not statistically significant. Figure \ref{fig:4b} shows that the only significant variable in accounting for the rural Han-Mongol wage differential is the lower average age of Mongols, which explains 5.2\% of the wage gap. 

\begin{table}[p]
\centering
\captionsetup{aboveskip=4pt, belowskip=6pt}
\caption{Blinder-Oaxaca Decomposition Results for Rural Area}
\label{tab:5}
\begin{threeparttable}
\setlength{\tabcolsep}{5pt}
\renewcommand{\arraystretch}{1.15}
\footnotesize
\begin{adjustbox}{width=\textwidth,center}
\begin{tabular}{@{}lcccccc@{}}
\\
  \toprule
\textbf{} &
  \textbf{} &
  \multicolumn{3}{c}{\textbf{Rural}} \\
  \cmidrule(lr){3-5}
 &
   &
  Han vs. Mongol &
  Han vs. Manchu &
  Manchu vs. Mongol \\
  \midrule
\multirow{2}{*}{Age} &
  Explained &
  \begin{tabular}[c]{@{}c@{}}0.052*\\ (0.025)\end{tabular} &
  \begin{tabular}[c]{@{}c@{}}-0.021\\ (0.017)\end{tabular} &
  \begin{tabular}[c]{@{}c@{}}-0.001\\ (0.056)\end{tabular} \\
 &
  Unexplained &
  \begin{tabular}[c]{@{}c@{}}0.882\\ (0.747)\end{tabular} &
  \begin{tabular}[c]{@{}c@{}}0.344\\ (0.546)\end{tabular} &
  \begin{tabular}[c]{@{}c@{}}0.611\\ (0.784)\end{tabular} \\
\multirow{2}{*}{Gender} &
  Explained &
  \begin{tabular}[c]{@{}c@{}}-0.019\\ (0.036)\end{tabular} &
  \begin{tabular}[c]{@{}c@{}}0.013\\ (0.027)\end{tabular} &
  \begin{tabular}[c]{@{}c@{}}-0.026\\ (0.038)\end{tabular} \\
 &
  Unexplained &
  \begin{tabular}[c]{@{}c@{}}-0.052\\ (0.333)\end{tabular} &
  \begin{tabular}[c]{@{}c@{}}-0.129\\ (0.196)\end{tabular} &
  \begin{tabular}[c]{@{}c@{}}0.071\\ (0.448)\end{tabular} \\
\multirow{2}{*}{Years of Schooling} &
  Explained &
  \begin{tabular}[c]{@{}c@{}}0.051\\ (0.033)\end{tabular} &
  \begin{tabular}[c]{@{}c@{}}-0.006\\ (0.014)\end{tabular} &
  \begin{tabular}[c]{@{}c@{}}0.096\\ (0.056)\end{tabular} \\
 &
  Unexplained &
  \begin{tabular}[c]{@{}c@{}}0.306\\ (0.527)\end{tabular} &
  \begin{tabular}[c]{@{}c@{}}0.430\\ (0.494)\end{tabular} &
  \begin{tabular}[c]{@{}c@{}}-0.164\\ (0.752)\end{tabular} \\
\multirow{2}{*}{Family Size} &
  Explained &
  \begin{tabular}[c]{@{}c@{}}0.001\\ (0.006)\end{tabular} &
  \begin{tabular}[c]{@{}c@{}}-0.001\\ (0.009)\end{tabular} &
  \begin{tabular}[c]{@{}c@{}}0.005\\ (0.017)\end{tabular} \\
 &
  Unexplained &
  \begin{tabular}[c]{@{}c@{}}-0.236\\ (0.582)\end{tabular} &
  \begin{tabular}[c]{@{}c@{}}-0.300\\ (0.282)\end{tabular} &
  \begin{tabular}[c]{@{}c@{}}0.062\\ (0.641)\end{tabular} \\
\multirow{2}{*}{Public/Gov Job} &
  Explained &
  \begin{tabular}[c]{@{}c@{}}0.009\\ (0.024)\end{tabular} &
  \begin{tabular}[c]{@{}c@{}}0.000\\ (0.004)\end{tabular} &
  \begin{tabular}[c]{@{}c@{}}0.016\\ (0.032)\end{tabular} \\
 &
  Unexplained &
  \begin{tabular}[c]{@{}c@{}}0.010\\ (0.145)\end{tabular} &
  \begin{tabular}[c]{@{}c@{}}-0.000\\ (0.032)\end{tabular} &
  \begin{tabular}[c]{@{}c@{}}0.003\\ (0.119)\end{tabular} \\
\multirow{2}{*}{Communist} &
  Explained &
  \begin{tabular}[c]{@{}c@{}}0.022\\ (0.019)\end{tabular} &
  \begin{tabular}[c]{@{}c@{}}-0.003\\ (0.006)\end{tabular} &
  \begin{tabular}[c]{@{}c@{}}0.017\\ (0.040)\end{tabular} \\
 &
  Unexplained &
  \begin{tabular}[c]{@{}c@{}}0.087\\ (0.119)\end{tabular} &
  \begin{tabular}[c]{@{}c@{}}-0.027\\ (0.019)\end{tabular} &
  \begin{tabular}[c]{@{}c@{}}0.123\\ (0.090)\end{tabular} \\
 &
   &
   &
   &
   \\
\multirow{3}{*}{Average Log Income Differences} &
  Total &
  -0.181 &
  0.085 &
  -0.267 \\
 &
  Explained &
  \begin{tabular}[c]{@{}c@{}}0.116\\ (0.068)\end{tabular} &
  \begin{tabular}[c]{@{}c@{}}-0.018\\ (0.037)\end{tabular} &
  \begin{tabular}[c]{@{}c@{}}0.107\\ (0.087)\end{tabular} \\
 &
  Unexplained &
  \begin{tabular}[c]{@{}c@{}}-0.297\\ (0.170)\end{tabular} &
  \begin{tabular}[c]{@{}c@{}}0.103\\ (0.107)\end{tabular} &
  \begin{tabular}[c]{@{}c@{}}-0.374*\\ (0.170)\end{tabular} \\
 &
   &
   &
   &
   \\
\multirow{3}{*}{Average Income Differences} &
  Total &
  1362.01 &
  11300.82 &
  -9938.81 \\
 &
  Explained &
  \begin{tabular}[c]{@{}c@{}}2376.70*\\ (1129.02)\end{tabular} &
  \begin{tabular}[c]{@{}c@{}}-547.45\\ (947.82)\end{tabular} &
  \begin{tabular}[c]{@{}c@{}}2417.48\\ (3183.25)\end{tabular} \\
 &
  Unexplained &
  \begin{tabular}[c]{@{}c@{}}-1014.70\\ (2933.08)\end{tabular} &
  \begin{tabular}[c]{@{}c@{}}11848.27\\ (7159.28)\end{tabular} &
  \begin{tabular}[c]{@{}c@{}}-12356.29*\\ (6393.91)\end{tabular}\\
  \bottomrule
\end{tabular}
\end{adjustbox}

\begin{tablenotes}[flushleft]\footnotesize
\item Significance codes: *** $p<0.001$, ** $p<0.01$, * $p<0.05$.
\item Standard errors in parentheses.
\item In group A vs B, results are showing the gap = B - A.
\end{tablenotes}
\end{threeparttable}
\end{table}

\subsubsection{Han-Manchu Gap}\label{subsubsec2}
In the urban sample, the average Manchus annual income is ¥10,973.46 lower than the average Han annual income. Akin to the urban Han-Mongol gap, ¥2,095.70 of the disparity is explained by Manchu’s lower participation in the CCP and the public sector according to Blinder-Oaxaca decomposition results. These two variables together contribute 4.4\% to the gap. Notice that here, education is not a significant variable explaining the wage gap. As seen in Table \ref{tab:1}, the average years of schooling are similar for Manchus and Hans, therefore educational difference is not large enough to account for the income difference. 

The direction of the differential is reversed in the rural case for Hans and Manchus, with the Manchus on average earning ¥11,300.82 more than the Hans. However, both the explained and unexplained parts of this gap are insignificant, implying that the rural Han-Manchu income gap is merely due to the large variations in wages.

\subsubsection{Manchu-Mongol Gap}\label{subsubsec3}
The urban Manchu-Mongol wage gap is ¥20,343.23, with Mongols having the higher average earnings. Here, the group differences in endowment variables explain ¥6,049.52, significant at 0.05 level. Specifically, Mongols’ higher average years of schooling account for 6.6\% of the higher income, and their higher probability of employment in the public sector contributes to 10.9\% of the wage gap. It is worth noticing that the unexplained wage gap is also significant, meaning that ¥14,292.71 of the higher income of Mongols remains unaccounted by any of the included variables. 

In rural areas, Manchus instead have an average annual income of ¥9,938.81 higher than Mongols. The Blinder-Oaxaca decomposition demonstrates that the explained part is not statistically significant, but the unexplained part is significant. This result demonstrates that, without unexplained discriminatory factors, Mongols should have the same earnings as Manchus. Discrimination lowers the average Mongol income by ¥12,356.29.

\subsection{Discussion}\label{subsec5.4}
Results from all three analyses suggest that while the “insider” Manchu group fares relatively similarly to the majority group Han, the “outsider” Mongol group suffers from an ethnic income penalty in rural regions after factoring in the effects of possible explanatory variables. In urban areas, this penalty vanishes because of increased participation in the public sector or governmental jobs and membership of the Chinese Communist Party of Mongols. Intuitively, the results are reasonable given the context of sampling from the Inner Mongolia autonomous region of Mongols. Under the urban setting, a preference for Mongols exists because affirmative action encourages firms to hire Mongols when they have similar qualifications as other candidates. Although the policy does not directly affect the hiring decisions of companies in the private sector, private firms benefit from a better reputation for having more Mongol employees. Also, for private firms based in Inner Mongolia, hiring Mongols who share similar culture, customs, and linguistic manners as the team would create a more harmonious working atmosphere, resulting in easier communication and increased efficiency. This serves as a possible explanation for the significantly high unexplained rural Manchu-Mongol wage gap. In the rural environment, discrimination is more prevalent because of the more intense racial prejudice against Mongols. With lower educational attainment levels and insufficient exposure to diversity, the stereotypical prejudice is entrenched in the rural culture. In addition, within the two provinces under examination, agricultural activities constitute the main economic pursuits in rural areas. Traditionally a nomadic group, Mongols are heavily involved in farming and pastoralism. The discrepancy between the wages for public or government jobs and agriculture-based jobs is very high, which effectively explains Mongols’ higher returns to public jobs in rural areas.

\section{Conclusion}\label{sec6}
In summary, this paper employs three statistical methods to analyze the income gap between an “outsider” minority group Mongols, an “insider” minority group Manchus, and the majority group Han in urban and rural regions in Inner Mongolia and Liaoning province in China. The dataset from the 2018 Chinese Household Income Project (CHIP) survey is used to perform the analyses. The findings reveal an unexplained rural Han-Mongol wage gap favoring the Han after considering group differences in objective factors related to income levels and a significantly positive correlation between involvement in the CCP and public sector occupation with income level in the urban areas. The Mongols in rural areas also show higher returns to holding a public sector job and CCP membership, implying the effectiveness of the ethnic preferential policy towards the “outsider” Mongol group. The results confirm trends concluded from previous studies and the initial hypothesis. 

The findings of this paper help evaluate the current ethnic policy in China and possibly have useful future policy implications. Although extrapolation of the results might not apply to all other ethnic minority groups in other regions of China, it is certain that affirmative actions, especially the promotion of minority preference in hiring, play a crucial role in creating a more ethnically equitable Chinese labor market. Aside from the intended channel of narrowing the wage gap by creating more employment opportunities for the Mongols in the government and public jobs, such policy also possibly associates a better reputation with hiring minorities in private firms. A possible way to further strengthen the contribution of the private sector to closing the gap is to establish preferential treatment policies for private firms that hire more minorities. While most current ethnic policy applies to all minority groups, the government should target the more recognizable minority groups to shrink the discrimination-based ethnic differences in income more efficiently. Specifically, the contemporary preferential hiring requirements are limited to the minority’s autonomous region; an extension of similar requirements to a wider range of regions would help the integration of “outsider” minorities in the labor markets with the Han and the “insider” minorities. 

\newpage

\bibliographystyle{plainnat}
\bibliography{bibliography}

\end{document}